\documentclass[a4paper,10pt]{scrartcl}
\usepackage[utf8]{inputenc}
\usepackage{graphicx,geometry,shapepar,tabularx}
\usepackage{amssymb}
\usepackage{ifpdf}
\usepackage{color}
\usepackage[all]{xy}
\usepackage{latexsym}
\newtheorem{theorem}{Theorem}[section]%\renewcommand{\theteorema}{}
\newtheorem{proposition}[theorem]{Proposition}%\renewcommand{\theproposicion}{}
\newtheorem{remark}[theorem]{Remark}%\renewcommand{\theobservacion}{}
\newtheorem{lemma}[theorem]{Lemma}%\renewcommand{\thelema}{}
\newtheorem{corolario}[theorem]{Corollary}%\renewcommand{\thelema}{}
\newtheorem{definition}[theorem]{Definition}%\renewcommand{\thedefinicion}{}
\newtheorem{example}[theorem]{Example}%\renewcommand{\thedefiniciones}{}

\title{Hausdorff separability of the boundaries \\ for spacetimes and sequential
spaces.}
\author{{\bf\large %L. Ak\'e??,
J.L. Flores$^*$,
J. Herrera$^\ddag$,
M. S\'anchez$^\dagger$}\\
{\it\small $^*$Departamento de \'Algebra, Geometr\'{\i}a y Topolog\'{\i}a,}\\
{\it \small Facultad de Ciencias, Universidad de M\'alaga,}\\
{\it\small Campus Teatinos, 29071 M\'alaga, Spain}\\
{\it \small $^\ddag$Instituto de Matem\'atica e Estat\'istica,}\\
{\it \small Universidade de S\~ao Paulo,}\\
{\it \small Rua do Mat\~ao, 1010, Cidade Universitaria S\~ao Paulo, Brazil}\\
{\it\small $^\dagger$Departamento de Geometr\'{\i}a y Topolog\'{\i}a,}\\
{\it\small Facultad de Ciencias, Universidad de Granada,}\\
{\it\small Avenida Fuentenueva s/n, 18071 Granada, Spain}}
\newcommand{\cvd}{\ \rule{0.5em}{0.5em}}
\newcommand{\Demo}{\noindent {\it Proof. }}
\newcommand{\tnew}{\tau_{chr}^*}
\newcommand{\lcrono}{L_{chr}}

\newcommand{\tcrono}{\tau_{chr}}
\newcommand{\lnew}{L_{chr}^{*}}
\newcommand{\lnewd}{L^{*}}
\newcommand{\tnewd}{\tau^*}
\newcommand{\taux}{\tau'}
\newcommand{\laux}{L'}
\newcommand{\eqref}[1]{(\ref{#1})}
\newcommand{\tadmisible}{{$T_2 $-admissible}}

\newcommand{\lnewm}{L_{chr}^{\,**}}
\newcommand{\tnewm}{\tau_{chr}^{**}}
\newcommand{\ltiD}{L\mid_{D}}

\newcommand{\crust}{minimally $D$-separating }
\newcommand{\crustL}{sequentially minimally $D$-separating}
\newcommand{\R}{\mathbb{R}}
\newcommand{\LL}{\mathbb{L}}
\newcommand{\N}{\mathbb{N}}
\newcommand{\tnewdl}{\tau^*_{Seq}}
\newcommand{\Aminl}{A$_{Min}^{Seq}$}
\newcommand{\Amin}{A$_{\hbox{\scriptsize Min}}$}
\newcommand{\Cambios}{}
\newcommand{\Asep}{A$_{\hbox{\scriptsize Sep}}$}
\newcommand{\ALMin}{$\tilde{\rm A}_{\hbox{\scriptsize Min}}^{\scriptsize Seq}$}
\newcommand{\AFin}{A$_{\hbox{\scriptsize Fin}}$}
\newcommand{\tauseq}{\tau_{Seq}}
\newcommand{\tnewL}{(\tau_{chr}^*)_{Seq}}
\newcommand{\AminM}{$\overline{A}_{\hbox{\scriptsize Min}}$}
\newcommand{\AminML}{$\overline{A}_{\hbox{\scriptsize Min}}^{\scriptsize Seq}$}

\begin{document}

\maketitle

\begin{abstract}
There are several ideal boundaries and completions in General
Relativity %(g-boundary, b-boundary, etc.)
 sharing the topological
property of being  {\em sequential}, i.e.,  determined by the
convergence of its sequences and, so,  by some limit operator $L$. As emphasized in a classical article by Geroch, Liang and Wald, some of them have the property,
commonly regarded as a  drawback, that there are points of the
spacetime $M$ non $T_1$-separated from points of the
boundary $\partial M$.

Here we show that this problem  can be solved
from a general topological viewpoint. In particular, there is a canonical minimum refinement of the topology in the completion $\overline{M}$
which T$_2$-separates the spacetime $M$ and its boundary $\partial M$  ---no matter the type of completion one chooses.
Moreover,
%we focus on
 % the process to define a topology in terms of a limit operator $L$ (which characterizes the possible convergence of its sequences)
  %, as commonly used in relativistic completions. This procedure does not make sense for any topological space but it does in the very general class of
 we analyze the case  of sequential  spaces and  show how the refined $T_2$-separating topology
 %for the completion
 can be constructed from a modification $L^*$ of the original limit operator $L$. Finally, we particularize this procedure to the case of the causal boundary and show how the separability of $M$ and  $\partial M$ can be introduced as an abstract axiom in its definition.

\end{abstract}
\begin{quote}
\end{quote}
\begin{quote}
	{\small\sl MSC:} {\small Primary 53C50, 83C75,\hspace{0.2cm}  Secondary: 54D55, 54A20.}
\end{quote}

\newpage

\tableofcontents

\section{Introduction}

A recurrent topic in the mathematical study of relativistic
spacetimes, is the definition of some types of ideal boundaries
which  encode relevant information about them. Being this a
natural practice from a mathematical viewpoint, there are further
physical motivations coming from the holographic principle and the
anti-de Sitter/conformal field theory correspondence.
There have been  quite a few   of proposals of boundaries
(including geodesic, bundle, abstract, conformal and causal
ones, see Section \ref{s4}), each one with its own
limitations. One of the most common problems appears from the
topological viewpoint, as typically one may find that some point of
the spacetime $M$ is not topologically well separated (i.e.,
non-Hausdorff separated or even non-T$_1$-separated) from some
point of the boundary $\partial M$. Such a property seems so
undesirable, that has been considered as a reason to reject {\em a
priori} many possible boundaries \cite{GLW}. The purpose of this
paper is to reconsider this question from a broad mathematical
perspective.

In a naive approach to the problem, one would like that a boundary
$\partial M$  satisfied: (i) it inherits the good topological
properties of the topology of the spacetime $M$, and (ii) it
records information of missing points, that is, it would satisfy, for example: if
a point of the spacetime is removed, the boundary will identify
this unequivocally and will allow to restore the missing point.  However, a closer look shows that these requirements would be too restrictive. About the requisite (i), notice first that one can find an open subset $D$
of Euclidean space $\R^n$ with, say, a fractal topological
boundary $\partial D\subset \R^n$. A worse contradiction with (i) appears when some of the commented
relativistic boundaries yield naturally a couple of non T$_2$-separated boundary points: this becomes a
natural consequence of the fact that there is no any distance
associated to a spacetime\footnote{When a distance exists on $M$
(as happens, for example, when $M$ is endowed with  a -positive
definite- Riemannian metric) one may expect that it will be
extensible to the boundary $\partial M$ and, thus, $\partial M$
would be Hausdorff. But recall that even a non-symmetric distance,
as the one associated to a non-reversible Finsler metric, may lose
properties when extended to the boundary. In fact,  the Cauchy
completion of a Finsler manifold may not be $T_1$, see
\cite[Section 3]{FHSst}. This is relevant for the case of spacetimes too, as Finsler metrics appear
naturally when the conformal class of simple classes of spacetimes
are considered \cite{CJS,FHSst}.}. About (ii), assume that one is considering a boundary
invariant by homotheties. Then one cannot
distinguish between removing a point and removing a closed ball in
Euclidean space, that is, between $\R^n\backslash\{0\}$ and
$\R^n\backslash \bar B(0,1)$. Thus, it would not be clear if such a
boundary should restore either a point or a  sphere. This shows that
(ii) may be also too restrictive; for example, it cannot be
expected for any conformally invariant boundary,  as some of the
relativistic ones.

Notice that, in certain sense, the bad
topological separation between some  $p\in M$ and some $Q\in
\partial M$, appears naturally when the requirements (i) and (ii)
fail: by the failure of the former, one can admit the bad
topological separation of two boundary points $P,Q\in \partial M$
but, because of the failure of the latter,   one might discover
that $P$ admits an interpretation as a missing point $p$ for some
new spacetime. Roughly, our answer to this problem will be the
following:

\begin{quote} {\em No matter how the completion $\overline{M} = M\cup \partial M$
is constructed, one can define a unique  minimal {\em refinement}
of the topology (i.e. a finer topology) in $\overline{M}$ which
will $T_2$-separate the points of the manifold $M$ and those of the
boundary $\partial M$. So (whenever the spacetime $M$ is clearly
identified in the completion $\overline{M}$), this refined
completion would not be rejected {\em a priori} by claiming
pathological separability properties. Moreover, the minimal
character of the refinement will preserve most of the desirable
properties of the original topology. }
\end{quote}

Once this aim is carried out, our next aim is motivated by the following observation.
Some of the previous topologies are defined in terms
of a limit operator $L$ which characterizes the convergence of the
sequences. Thus, it is interesting to study especially this case and, concretely, how to construct a Hausdorff separating topology by modifying directly $L$.
Mathematically, the spaces whose topology can be characterized by
such an operator
 are the {\em sequential spaces}. This is a vast class of spaces
(it includes all the first countable ones, in particular the
metric spaces, see Section \ref{s}). So, we will also carry out  a
study of independent mathematical interest about the refinement
process for sequential spaces endowed with a limit operator.

As a last aim, this
study will be applied specifically to the causal boundary (c-boundary for short in the remainder)  of
spacetimes. The reason is twofold. On the one hand, this boundary
has been widely studied recently, because it seems the unique
intrinsic and general  alternative to the Penrose conformal
boundary (the latter is commonly used in Mathematical Relativity,
but its existence is ensured only in quite particular
cases). On the other hand, the redefinition of the c-boundary in
\cite{FHS} left open the possibility of new refinements of such a
notion, by imposing additional properties to the boundary which
might be judged as  desirable a priori. The property of separation
studied here is a good example of this possibility.

This article is organized as follows. In Section \ref{s2} we
recall some general features of sequential spaces, and
prove some first  properties about limit operators (Subsection \ref{s2.1}, specially Propositions
\ref{p} and \ref{lem2}).
The following subtlety is emphasized. A limit operator
$L$ for a topology $\tau$ is {\em full} or {\em of first
order} when $L$ provides directly all the limits of all the
convergent sequences for $\tau$; in this case, it is univocally determined and can be denoted by
$L_\tau$. Otherwise,  $L$ determines only some limits of
convergent sequences, which are enough to determine the topology. Even more, one can wonder if $L$ is of second order (roughly, the iteration of $L$ twice is enough to determine $L_\tau$)  or of other higher orders; this is briefly explained in Subsection \ref{s2.2}. Even though any topology (sequential or not) admits  a unique first order $L_\tau$,  it is interesting to consider also the case of non-first order limit operators
 because: (a) the topology of some relativistic boundaries is defined by imposing that {\em naturally, some sequences must converge}; however,
such a convergence  may imply some other (possibly undesired) limits, and
(b)~in principle, simple modifications of $L$
 (as, eventually, those required for our refinement of the topologies) might give limit operators that
 are not of first order, even if $L$ was.

Section \ref{s}. In Subsection \ref{s3.1},  the problem of
$T_2$-separating a domain $D$ of its complement $X\backslash D$, in an arbitrary topological space $(X,\tau)$ (see Defn. \ref{dcrust}), is
considered. A (minimum)  unique explicit topology $\tau^{*}$
is shown to solve the problem (Theorem \ref{prop2}). In Subsection \ref{s3.2}, we
consider the problem for the special case of sequential spaces.
Recall that for these spaces, it is natural to wonder about a  separating topology which is minimum {\em among the sequential ones}.
 The existence of
a unique sequential topology $\tau^*_{Seq}$ solving this problem  of $T_2$-separation is also obtained (Theorem \ref{t1}). Finally, in Subsection \ref{sec3.3} we analyze the role of the chosen limit operator $L$ for the  case of sequential topologies. Concretely, the domain $D$
allows to define naturally a modification $L^{\ast}$ of the
original operator $L$ (see formula \eqref{def2}). When $L$ is of first order, the topology
generated by $L^{\ast}$ can be identified with the topology previously
obtained (see Theorem~\ref{prop1} and Figure~\ref{figresumen}; notice that $L^*$ is then necessarily of first order too); otherwise, some
other properties still hold (Proposition \ref{prop1'}).

Section \ref{s4}. In Subsection \ref{s4.1} we review briefly the most common
boundaries for spacetimes, and explain how previous results can be
applied to ensure the Hausdorff separability between points of $M$
and $\partial M$. Special care is put in the discussion of a
general example by Geroch, Liang and Wald \cite{GLW} which shows
that, under general hypotheses, the g-boundary and other boundaries introduced in
General Relativity will have a pair of non-$T_1$ related points,
one of them in the spacetime and the other in the boundary. Our
previous results allow to circumvent this problem and suggest that the
g-boundary should be redefined. In the remainder we
focus in the case of the c-boundary, whose topology (the
chronological topology $\tau_{chr}$) is defined as the one derived from some
limit operator $\lcrono$. In Subsection \ref{s4.2}, a short review of the causal
completion is provided. In Subsection \ref{s4.3} we check that  the good properties of this completion
(previously summarized in Theorem \ref{theo1}) are maintained when one uses both,
the corresponding separating topology $\tau_{chr}^*$ and the sequentially
separating one $(\tau_{chr}^*)_{Seq}$ (Theorem \ref{theo2}).

Section \ref{s5}. Here, we
analyze the process to define  the c-boundary from  very
general admissibility properties, according to \cite{FHS}. In Subsection \ref{s5.1}, the admissibility conditions in \cite{FHS} are
revisited, obtaining an extension of the results in that reference
which include non-necessarily sequential topologies (Theorem
\ref{theo5}). In Subsection \ref{s5.2}, the following aim is achieved.
Assume that  the
Hausdorff separability between points of $M$ and $\partial M$ is
incorporated as one of these admissibility properties that characterize the c-boundary.
Then the corresponding topology becomes equal to the chronological
topology refined in the general $D$-separating way explained in Section \ref{s}, under the hypothesis that $\lcrono$ is of first order and even under less restrictive hypothesis which exhibit the accuracy of the approach
 (see Theorem \ref{t}, the discussion in Remark
\ref{g} and the summarizing Figure \ref{figresumen2}).
% the topology generated by a
%natural operator $\lcrono^*$ is naturally obtained.
%In particular, when
%the operator $\lcrono$ is of first order, this topology agrees
%with the corresponding topology $\tau^*$ (see Figure 2??).

The article finishes with an Appendix containing examples which show that
most of the subtleties suggested by our procedures can occur effectively.

\section{Limit operators and sequential spaces}\label{s2}
In this section we develop some properties about limit operators
and sequential spaces. There is a well established literature on
the later, see  \cite{Franklin, Goreham}  and references therein
for general background.

\subsection{General properties}\label{s2.1}

Let $X$ be an arbitrary set, let $S(X)$ the set of all sequences
in $X$ and  ${\cal P}(X)$ the set of parts of $X$. Following
\cite[Sect. 3.6]{FHS}, \cite[Sect.5.2.2]{FHSst} we consider:

\begin{definition}\label{def3}
A map $L:S(X)\rightarrow {\cal P}(X)$ is a {\em limit operator} if
it satisfies the following {\em compatibility condition for
subsequences}:
\begin{equation}\label{***}
L(\sigma)\subset L(\kappa),\quad \hbox{for any $\sigma,\kappa\in
S(X)$ with $\kappa\subset\sigma$.}
\end{equation}
In this case, the {\em topology derived
%\footnote{M: creo que
%usabamos derived}
from $L$} is the topology $\tau_L$ whose closed sets are those
subsets $C$ of $X$ such that $L(\sigma)\subset C$ for any sequence
$\sigma\subset C$.
\end{definition}
In fact, the compatibility condition for subsequences allows to
prove easily that the so-defined closed subsets satisfy the axioms
for a topology.

\begin{remark}\label{rcoherent} {\em
(1) We can assume with no loss of generality that any limit operator is {\em coherent} in the sense that, for any $x\in X$ one has $x\in L(\{x_n=x\})$. This will be assumed explicitly  in the case of limit operators of $k$-th order below.

(2) If a topology $\tau$ on $X$ has been prescribed, one can
define a unique {\em associated limit operator} $L_\tau$ which
gives directly the convergence of sequences in $(X,\tau)$ (i.e.,
such that the converse of formula \eqref{eq1} also holds), namely:
\begin{equation}\label{eq11} L_\tau:S(X)\rightarrow {\cal P}(X),\quad L_\tau (\sigma):=\{p\in X:\; \sigma\rightarrow p\;\hbox{with
$\tau$}\}.\end{equation}

(3) For an arbitrary limit operator $L$, the derived topology satisfies clearly the following
implication:
\begin{equation}\label{eq1}
p\in L(\sigma)\;\Longrightarrow\; \sigma\rightarrow p \hbox{ with
the topology $\tau_{L}$}.
\end{equation}
However, the converse does not hold in general
and we will be
also interested in this possibility.
A trivial (non-coherent) example
is
 just the limit operator $L(\sigma)=\emptyset$ for
all $\sigma\in S(X)$, whose derived topology is the discrete one.
A coherent example can be constructed  as follows. Let  $(X,\tau)$ be any sequential topological space with a convergent sequence  $\sigma =\{x_n\}\rightarrow x_\infty$ such that no subsequence of $\sigma$ is equal to $\sigma$, except $\sigma$ itself (for example, the elements of $\sigma$ could be all distinct as in the case $(X,\tau)=\R, \sigma=\{1/n\}, x_\infty=0$). Let $L_\tau$ be the associated limit operator and
 define a new limit operator $L$ such that  $L(\tilde\sigma)=\emptyset$, where $\tilde \sigma$ is any sequence that contains $\sigma$ as a subsequence, and $L(\mu) =L_\tau(\mu)$ for any other sequence $\mu$. As $\mu$ can be chosen equal to any subsequence of $\sigma$ (except $\sigma$ itself), this implies that, in any case, $\sigma$ will converge to $x_\infty$ with $\tau_{L'}$ and, then, $\tau_{L'}=\tau$.

 Obviously, these examples correspond to limit operators where $L$ is  regarded as empty on a sequence in a highly arbitrary way. Nevertheless,  one may find naturally the following situation in the case of non-Hausdorff topologies: the limit operator yields a infinite set of limits, and these limits will yield naturally more limits. This will lead to the notion of $k$-th limit operator, to be studied below.

}\end{remark}
The  considerations above motivate the following definition.
\begin{definition}
%\label{def6}
Let $D$ be a subset of $X$. We will say that a limit operator $L$
on $X$ is of {\em first order on $D$} if, for any sequence
$\sigma\subset D$ and any $p\in D$:
\[
p\in L(\sigma) \iff \sigma\rightarrow p\;\;\hbox{with $\tau_{L}$}.
%\quad\hbox{for any sequence $\sigma\subset D$ and any $p\in D$.}
\]
When $L$ is of first order on $D=X$ (i.e. $L=L_{\tau_L}$) then  we
say simply  that $L$ is of {\em first order}.
\end{definition}
As a first simple property to be used in the remainder, we have
the following.
\begin{lemma}\label{lem3.3a} Let $L, L'$ be two limit operators on $X$ with derived topologies $\tau_L %(\equiv \tau)
$ and $\tau_{\laux}$ resp.  If
\begin{equation}\label{*}\laux(\sigma)\subset L(\sigma)\qquad\hbox{for any sequence $\sigma\subset X$,}\end{equation}
then $\tau_L \subset \tau_{\laux}$. Moreover, when $L$ is of first
order, the converse  also holds.
\end{lemma}

{\Demo} Let $C$ be a closed set for $\tau_L$, that is,
$L(\sigma)\subset C$ for any sequence $\sigma\subset C$.
%(Defn.
%\ref{def3}).
As $\laux(\sigma)\subset L(\sigma)\subset C$, we deduce that $C$
is also closed for $\tau_{\laux}$, and so, $\tau_L \subset
\tau_{\laux}$.

 For the converse, assume that $L$ is of first order
and $q\in \laux(\sigma)$ for some sequence $\sigma\subset X$.
Then, $\sigma\rightarrow q$ for $\tau_{\laux}$, and, by the
inclusion $\tau_L\subset \tau_{\laux}$, this limit also holds for
$\tau_L$. As $L$ is of first order, necessarily $q\in L(\sigma)$,
as required. \cvd

 Recall that a topological space $(X,\tau)$
is called {\em sequential} if any {\em sequentially closed set}
(i.e. a set which contains all the limits of any sequence
contained in it) is also a closed set. Note that the converse is
true in any topological space.
\begin{proposition} \label{p} Let $X$ be a set:

(a) For any limit operator $L$ on $X$, its derived topology
$\tau_L$ is sequential.

(b) For any topology $\tau$ on $X$, its associated limit operator
$L_\tau$ (see formula \eqref{eq11}) has a derived topology
$\tauseq :=\tau_{L_\tau}$, which is the coarsest one among the sequential
topologies containing $\tau$. Moreover, $L_\tau$ is a limit
operator of first order for $\tauseq$.

(c) A topology $\tau$ on $X$ is sequential if and only if
$\tau=\tauseq$.

%\smallskip

%\noindent
\end{proposition}
{\Demo} (a) is straightforward from the definitions.

For (b), the inclusion $\tau\subset\tauseq$ follows from
the fact that closed subsets are sequentially closed for any
topology. Let $\tau'$ be another sequential topology with $\tau\subset\tau'$. If $\sigma$ converges to $p$ with $\tau'$, then $\sigma\rightarrow p$ with $\tau$, which implies $p\in L_{\tau}(\sigma)$. Hence, sequentially closed sets for $\tauseq$ are also sequentially closed for $\tau'$, and so, $\tauseq\subset\tau'$.
%Let $\tau_{L'}$ another sequential topology with
%$\tau\subset\tau_{L'}$. If $p\in L'(\sigma)$ then $\sigma$
%converges to $p$ with $\tau_{L'}$, and thus, also with $\tau$, which
%implies $p\in L_{\tau}(\sigma)$. Hence, $L'(\sigma)\subset
%L_{\tau}(\sigma)$ for any sequence $\sigma$ and, from Lemma
%\ref{lem3.3a}, $\tau_{L_{\tau}}\subset \tau_{L'}$.
For the first
order character of $L_{\tau}$, assume that $\sigma$ converges to
$p$ with $\tauseq$. Since $\tau\subset \tauseq$,
the sequence $\sigma$ must converge to $p$ with $\tau$, and thus,
$p\in L_{\tau}(\sigma)$.

Finally, the right implication of (c) follows from the assumption that sequentially closed for
$\tau$ implies closed (to the left, use (a)).
% Finally, the last assertion follows from the previous ones.
 \cvd

\begin{remark}\label{r2.5} {\em
Summing up, a topological space $(X,\tau)$ is sequential if and
only if $\tau=\tau_{L}$ for some limit operator $L$ on $X$, which
can be chosen of first order (and, then, equal to $L_\tau$). From
the previous properties and the known ones about sequential spaces
(see \cite{Goreham}, especially its figures 1.1 and 1.3) the
following observations are in order:

(1) The map $\tau \mapsto L_\tau$ from the set  $\mathcal{T}(X)$
of all the topologies on $X$ to the set of all the first order
limit operators on $X$, is an onto map (as $L=L_{\tau_{L}}$). Nevertheless, it is not one to
one, as any non-sequential topology $\tau$ has associated the
limit operator $L_{\tau}$ equal to the limit operator for the
sequential topology $\tauseq$. In fact, one has then an
onto map
$$
\mathcal{T}(X)\rightarrow \mathcal{T}_{\scriptsize{\rm Seq}}(X),
\quad \quad \tau \mapsto  \tauseq
$$
where  $\mathcal{T}_{\scriptsize{\rm Seq}}(X)$ is the set of all
the sequential topologies on $X$.

(2) Sequential spaces generalize Fr\'echet-Uryshon spaces (namely,
the closure of a set consists of the limits of all the sequences
in that set) and, then, first countable spaces, which include all
the metrizable ones; however, sequential spaces can be regarded as
quotients of metrizable spaces.

 If $X$ is a sequential space, the continuity of a map $f:X\rightarrow Y$ into a topological space $Y$ can be characterized by the preservation of the limits of sequences. Nevertheless, the closure of a set $A\subset X$ may contain {\em strictly} the  set of all the limits of sequences in $A$ (in contrast  with Fr\'echet-Uryshon),
 the uniqueness of the limits of sequences does not imply Hausdorffness (in contrast with first countability), and  sequential compactness (i.e., the property that any sequence admits a  convergent subsequence) plus Hausdorffness and first countability, do not imply compactness (in contrast with metrizability). Nevertheless, for sequential spaces,
 sequential compactness becomes equivalent to countable compactness (see for example \cite[Proposition 3.2]{Goreham}) and, as the latter is weaker than compactness,  {\em compact sequential spaces are sequentially compact}.
%\footnote{M: Even more, compactness and sequential compactness are
%equivalent for paracompact sequential spaces \cite[Corollary
%6.10]{Goreham}.}
%Si la completaci'on de Busemannn fuera
%paracompacta (recordemos que era siempre secuencial \cite[Theorem
%5.28]{FHSst}) ser'ia tambi'en una compactificaci'on adem'as de una
%compactificaci'on secuencial (eso ser'ia un buen ejercicio para
%Luis y, si lo demuestra, o encuentra un contraejemplo, lo podria
%publicar e'l  en los Proc AMS casi con seguridad).
}\end{remark}
The property of being sequential is not inherited by arbitrary
subspaces, but it is inherited by open (or closed) subspaces
\cite[Prop. 1.9]{Franklin}. More sharply, the limit operators can
be inherited by open subspaces, in the following sense.
\begin{proposition}\label{lem2}
Let $D$ be an open subset of a sequential space $(X,\tau_L)$. The
topology $\tau_{L}\mid_{D}$ induced on $D$ by $\tau_L$ coincides
with the topology $\tau_{L\mid_{D}}$ derived from the {\em
restricted limit operator} $\ltiD$ given by
\[
\ltiD(\sigma):= L(\sigma)\cap D,\qquad\hbox{for any sequence
$\sigma\subset D$.}
\]
\end{proposition}
{\Demo} It is straightforward to check that $\ltiD$ is a limit
operator if so is $L$. So, we  focus on the identification between
$\tau_{L}\mid_{D}$ and $\tau_{L\mid_{D}}$.

In order to prove that $\tau_{L}\mid_D\subset \tau_{L\mid_{D}}$,
let $C$ be a closed set of $\tau_{L}\mid_D$. Then, $C=C'\cap D$
with $C'$ a closed set of $\tau_L$. Let $\sigma$ be any sequence
in $C$. As $\sigma\subset C\subset C'$ and $C'$ is closed for
$\tau_L$, necessarily $L(\sigma)\subset C'$. Therefore,
\[
L\mid_{D}(\sigma)=L(\sigma)\cap D\subset C'\cap D=C,\;\;\hbox{and
thus,}\;\; \hbox{$C$ is closed for $\tau_{L\mid_{D}}$.}
\]
For the inclusion $\tau_{L\mid_{D}}\subset \tau_{L}\mid_D$, let
$B\subset D$ an open set of $\tau_{L\mid_{D}}$. It is enough to
prove that $B$ is also open for $\tau_{L}$. So, assume by
contradiction that  $X\setminus B$ is not closed for $\tau_{L}$,
and thus, there exist a sequence $\sigma\subset X\setminus B$ and
a point $p\in B\subset D$ such that $p\in L(\sigma)$. As $D$ is
open for $\tau_L$, $\sigma\subset D$ eventually. In particular,
$p\in L(\sigma)\cap D=L\mid_{D}(\sigma)$, and from (\ref{eq1}),
$\sigma$ converges to $p$ with $\tau_{L\mid_{D}}$. But $B$ is open
for $\tau_{L\mid_{D}}$, hence $\sigma$ eventually intersects $B$,
which is a contradiction. \cvd

%\begin{definition}
%
%The topology $\tau_{L}$ is said to be of {\em first order} when
%
%\begin{equation}\label{eq3}
%x\in L(x_n) \iff \{x_n\}\rightarrow x \hbox{ with the topology $\tau_{L}$}.
%\end{equation}

%\end{definition}

%\input secseparable.tex

\subsection{Higher order limit operators}\label{s2.2}

When a limit operator $L$ is not of first order, $L(\sigma)$ does not provide all the topological limits of some sequence $\sigma\subset X$, that is, $L(\sigma)\varsubsetneq L_{\tau_L}(\sigma)$.
Nevertheless, in view of the discussion in Remark \ref{rcoherent} (as well as Example \ref{eluis} below), it is natural to wonder when any point in $L_{\tau_L}(\sigma)\setminus L(\sigma)(\neq\emptyset)$ can be obtained by iterating successively the limit operator $L$. In order to formalize this idea, we will assume that $L$ is a coherent limit operator (Remark \ref{rcoherent} (1)) and consider the following transfinite definition:
\begin{equation}\label{aux2}
L^1(\sigma):=L(\sigma),\qquad L^{i}(\sigma):=\{p\in X:\; p\in L(\{p_n\}),\;\;\{p_n\}\subset \cup_{j<i}L^j(\sigma)\}\quad\hbox{for any ordinal $i\geq 2$}.
\end{equation}
Note that if $p\in L^{k}(\sigma)$, then $p\in L(\{p_n=p\}_n)$ with $\{p_n\}_n\subset L^{k}(\sigma)\subset\cup_{j<i}L^j(\sigma)$ for any $i\geq k$, and thus, $p\in L^{i}(\sigma)$. Therefore, the following property holds:
\begin{equation}\label{cita}
L^{k}(\sigma)\subset L^{i}(\sigma)\qquad\hbox{for any sequence $\sigma$ and any ordinals $k\leq i$.}
\end{equation}

%Then, for a fixed sequence $\sigma$, we have a sequence of nested sets $\{L^{i}(\sigma)\}_i$ which limit motivates the following definition:
%
%\[
%L^\infty(\sigma)=\cup_{i=1}^{\infty} L^i(\sigma).
%\]
%
%Now, we are in conditions to prove the following proposition:
\begin{proposition}\label{aux1} The inclusion $L^{i}(\sigma)\subset L_{\tau_L}(\sigma)$ holds for any $\sigma\subset X$ and any ordinal $i$.
\end{proposition}
{\it Proof.} One needs to show that if $p\in L^i(\sigma)$ then $\sigma$ converges to $p$ with the topology $\tau_L$, i.e., if $\sigma=\{p_n\}$ and $p\in U$ for some open set $U$ of $\tau_L$, there exists $n_0$ such that $p_n\in U$ for all $n\geq n_0$.

The proof will follow by transfinite induction. For $i=1$, the result is known from (2). So, assume that it is true for all $j<i$, and let us prove it for $i$. From the definition of $L^i(\sigma)$, there exists a sequence $\{q_{n}\}\subset \cup_{j<i}L^j(\sigma)$ such that $p\in L(\{q_n\})$. From (2) the sequence $\{q_n\}$ converges to $p$ with the topology $\tau_L$, and so, there exists some $n'_0$ such that $q_{n'_0}\in U$. Since $q_{n'_0}\in \cup_{j<i}L^j(\sigma)$, necessarily $q_{n'_0}\in L^j(\sigma)$ for some $j<i$, and the hypothesis of induction ensures that $\sigma$ converges to $q_{n'_0}$ with $\tau_L$. Recalling that $q_{n'_0}\in U$, there exists $n_0$ such that $p_n\in U$ for all $n\geq n_0$. Hence, $\sigma$ converges to $p$ with $\tau_L$. \cvd

Motivated by the notion of first order limit operator, we now establish the following definition.
\begin{definition}
A coherent limit operator $L$ is of $k$-th {\em order} for some ordinal $k$ if
\[
L_{\tau_L}(\sigma)=L^{k}(\sigma),\qquad\hbox{but}\quad L_{\tau_L}(\sigma)\neq L^{j}(\sigma)\quad\hbox{for any $j<k$}.
\]
\end{definition}
\noindent Recall that the coherent limit example constructed in Remark \ref{rcoherent}(3) is not of $k$-th order for any ordinal. As a direct consequence of (\ref{cita}), Proposition \ref{aux1} and this definition, we also deduce:
\begin{corolario}\label{cc} If a coherent limit operator $L$ is of $k$-th order then $L^k(\sigma)=L^{i}(\sigma)$ for any $\sigma\subset X$ and any ordinal $i\geq k$.
\end{corolario}

%\begin{remark} {\em It is important to recall that in previous definition we include the case of infinite order. As we will see, the main result of this appendix will remain valid even on such a case.}
%\end{remark}

\section{T$_2$-separating boundaries in the general and sequential cases}\label{s}

\subsection{Separating for an arbitrary topology} \label{s3.1}

In this section we consider the following problem. Let $(X,\tau)$
be a topological space, and let $D\subset X$ be an open subset
with some good topological properties, namely, Hausdorff and
locally compact. Typically, $D$ will be a Lorentzian manifold and
$X$ some topological completion, so that $D$ is dense in $X$.
However, we do not impose a priori the density, allowing
$X\backslash D$ to be a bigger ``crust''. Now, assume that there
is some pair of points $p\in D$, $q\in X\setminus D$ that are not
Hausdorff separated, and we look for a ``minimal refinement of
the topology around $\partial D$'' in order to ensure that such
pairs are $T_2$-separated. Formally, we are looking for a new
topology $\tnewd$ which satisfies the following requirements.

\begin{definition}\label{dcrust} Let $(X,\tau)$ be
 a topological space, and let $D\subset X$ be an   open subset which is Hausdorff and
locally compact. A  topology $\tnewd$ is {\em \crust} if it
satisfies:

\begin{itemize}
\item[(\AFin)] {\em Refinement of $\tau$}: the topology $\tnewd$ is  finer than $\tau$, i.e.
%\begin{equation}\label{e6'}
$\tau\subset \tnewd.$ %\quad \quad  .
%\end{equation}

\item[(\Asep)] {\em $T_2$-separability of
points of $D$ and $X\backslash D$}: for any $p\in D$, $q\in
X\setminus D$, there exist $U\in\tau$ %(\subset\tnewd)$
and $V\in\tnewd$ such that
%\begin{equation}\label{**'}ç
$p\in U,\; q\in V$ and %\;\hbox{and}\;\;
$U\cap V=\emptyset$.
%\end{equation}

\item[(\Amin)]{\em Minimality}:  $\tnewd$
is a minimal topology among those satisfying (\AFin) and
(\Asep), i.e. no other topology satisfying conditions (\AFin) and (\Asep) is
strictly coarser than $\tnewd$.
\end{itemize}
\end{definition}

\begin{remark} {\em (1) From property (\AFin), the subset $D$ will be also open for $\tnewd$.

(2) The condition of minimality (\Amin) is
essential in order to avoid trivial topologies. For example, if we
considered as $\tnewd$ the one generated by $\tau$ and the crust
$X\backslash D$ as a subbasis, then $\tnewd$ would satisfy (\AFin)
and (\Asep), but at the cost of a
separability which would forbid any point in the $\tau$-boundary
$\partial D$ to be $\tnewd$-continuously reachable from $D$ (i.e.,
the $\tnewd$-boundary of $D$ would be the empty set); a similar
drawback would happen if one considered the topology generated  by
$\tau$ and the sets $\{q\}$ for every point $q\in X\setminus D$
which is not T$_2$-separated from some point in $D$.

%Condition (A$_{\hbox{Min}}$) will ensure that if $D$ was
%$\tau$-dense in $X$ then it is also $\tnewd$-dense,  see Theorem
%\ref{prop2} (ii).\footnote{OJO: Esto habra que suprimirlo ahora.}

(3) There exists a small asymmetry in the condition
of
$T_2$-separability (\Asep), as the separating
neighborhood $U$ is not only required to belong to $\tnewd$ but
also to $\tau$. Notice that only $D$ is assumed to be a
topological subspace with additional  nice properties
(Hausdorfness, local compactness). This asymmetry in
(\Asep), plus the minimality in
(\Amin), will ensure that the topology
$\tnewd$ preserves most of the original properties of
$(X,\tau)$. Among them,  $\tau$ and $\tnewd$ will induce the same
topology on $D$ (see Theorem \ref{prop2} below). If the asymmetry
were not imposed, $\tau|_D \neq \tnewd|_D$ may occur (Example
\ref{ex0}). }
\end{remark}
Now, we can state the following general result for topological
spaces.
\begin{theorem}\label{prop2}
Let $(X,\tau)$ be a topological space and $D\subset X$ be a
Hausdorff locally compact open subset of $X$. The topology
$\tnewd$ generated by the subbasis
\begin{equation}\label{e9}
{\cal S}=\tau\cup \{X\setminus K:\;\; K\;\,\hbox{is a compact
subset of D}\}
\end{equation}
is the unique topology which satisfies the properties (\AFin),
(\Asep) and (\Amin)
above, i.e., the unique \crust topology.
%, i.e.,  the coarsest
%topology among the ones containing $\tau$ and T$_2$-separating the
%points in $D$ from the points in $X\setminus D$.
Moreover, the restrictions of  $\tnewd$ and $\tau$ on $D$
coincide.
%and both, the interior\footnote{M: modificado el enunciado, pero
%no la demostracion en esa cuestion; la demostracion  habria que
%completarla y explicarla un poco mejor (si' he cambiado alguna
%inconsistencia en ella.} of $X\setminus D$ are equal for $\tnewd$
%and $\tau$, and the restriction of both topologies to $ {\rm
%int}(X\setminus D)$ also coincide.

\end{theorem}
\Demo Clearly, $\tnewd$ satisfies (\AFin). In order to prove
(\Asep), consider $p\in D$ and $q\in
X\setminus D$. Let $K$ be a compact neighborhood of $p$ in $D$ and
$U$ its interior. From the definition of the subbasis ${\cal S}$
in (\ref{e9}), $V:=X\setminus K$ is open for $\tnewd$ and, so, $U$
and $V$ are the required open sets.

In order to prove (\Amin) and the uniqueness, we will show that $\tau^*\subset\taux$ for any other topology $\taux$ satisfying (\AFin) and
(\Asep). Since $\tau\subset \taux$, it suffices to show that $X\setminus K\in
\taux$ for any compact subset $K$ of $D$. So, let us prove that
$X\setminus K$ is a neighborhood of any $q\in X\setminus K$. From
the property (\Asep) and the Hausdorffness
of $D$, for every $p\in K$ there exists $U_p\in\tau\subset\taux$
and $V_p\in\taux$ such that $p\in U_p$, $q\in V_p$ and $U_p\cap
V_p =\emptyset$. Since $K$ is compact, there exists
$\{U_{p_i}\}_{i=1}^n\subset \{U_{p}\}_{p\in K}$ such that
$K\subset \cup_{i=1}^{n}U_{p_i}$. Then,
$V:=\cap_{i=1}^{n}V_{p_i}\in\taux$ satisfies $q\in V$ and $V\cap
K\subset V\cap (\cup_{i=1}^{n}U_{p_i})=\emptyset$. Therefore,
$X\setminus K$ is a neighborhood of $q$, as required.

For the last assertion, it suffices to show that $A\cap D\in\tau$
for any $A\in {\cal S}$ (recall that ${\cal S}$ is subbasis of
$\tau^*$). So, assume that $A=X\setminus K$, where $K$ is a
compact subset of $D$ (if $A\in\tau$, the conclusion follows
trivially from $D\in\tau$). Since $D$ is Hausdorff, $K$ is closed
in $D$, and thus, $D\setminus K (=A\cap D) \in\tau$, as required.
\cvd
%For (ii), let us assume by
%contradiction that there is a point $x$ which is in the
%$\tnewd$-interior but not in the $\tau$-interior of $X\setminus
%D$. This is only possible if $x$ belongs to some $U\in
%\tnewd\setminus \tau$ with $U\subset X\setminus D$. But, in this
%case, we can assume $U=X\setminus K$ for some compact subset
%$K\subset D$, and so, $U\cap D=D\setminus K\neq\emptyset$, which
%is a contradiction.
%Finally, the identity $\tnewd\mid_{{\rm int}(X\setminus
%D)}=\tau\mid_{{\rm int}(X\setminus D)}$ is a consequence of both,
%the inclusion $\tau\subset \tnewd$ and the fact that any open set
%$U=X\setminus K\in\tnewd$, for some compact set $K\subset D$,
%satisfies $U\cap (X\setminus D)=X\setminus D$. \cvd

%
%and the compactness of $K$, there exists two finite families
%$\{U_{i}\}_{i=1}^{n}$ and $\{V_{i}\}_{i=1}^{n}$ of open subsets of
%$\tau$ and $\tau_D$, resp., such that $K\subset
%U:=\cup_{i=1}^{n}U_{i}$, $p\in V:=\cap_{i=1}^{n}V_{i}$ and $U\cap
%V=\emptyset$. In particular, $p\in V\subset X\setminus K$, and
%thus, $X\setminus K$ is a neighborhood of $p$. \cvd

\subsection{Separating among sequential topologies}\label{s3.2}

Next, we consider the same problem as before but restricting our
attention to sequential spaces. To do that, first we have to
introduce the following adapted version of Definition
\ref{dcrust}:
\begin{definition}
Let $(X, \tau (=\tau_{L}))$ be a sequential topological space, and
let $D\subset X$ be an open subset which is Hausdorff and locally
compact. A topology $\tnewdl$ is {\em {\crustL}} if it satisfies
(\AFin) and (\Asep) from Definition
\ref{dcrust}, and the following minimality condition:
\begin{itemize}
\item[({\Aminl})] {\em Sequential minimality:} $\tnewdl$ is a
minimal topology among the {\em sequential ones} satisfying
(\AFin) and (\Asep).
\end{itemize}
\end{definition}
Now, we can state the following result for sequential topological spaces:
\begin{theorem}\label{t1}
Let $(X,\tau (=\tau_{L}))$ be a sequential topological space, and
let $D\subset X$ be a Hausdorff locally compact open subset of $X$. The topology $(\tau^*)_{Seq}$ (where $\tau^*$ is the unique {\crust}
topology according to Theorem \ref{prop2} and the notation introduced in Proposition \ref{p}(b) is used), is the unique {\em
{\crustL}} topology, that is. $\tnewdl$ exists,  it satisfies ({\Aminl}) as an unique minimum 
 and $(\tau^*)_{Seq}=\tnewdl$. Moreover, the restrictions of $\tnewdl$ and $\tau$ coincide on $D$.
\end{theorem}
{\Demo} From Proposition \ref{p} (b), the topology
$(\tau^*)_{Seq}$ is the coarsest topology among the sequential
ones containing $\tau^*$. So, $(\tau^*)_{Seq}$ clearly
satisfies conditions (\AFin) and (\Asep),
as these conditions are already satisfied by $\tau^*$. Moreover,
any other sequential topology $\tau'$ satisfying (\AFin) and
(\Asep), must contain $\tau^*$. So, from
the coarsest character of $(\tau^*)_{Seq}$, the topology
$\tau'$ must contain $(\tau^*)_{Seq}$. Therefore, we deduce
both, $(\tau^*)_{Seq}$ satisfies condition ({\Aminl}) and it is
unique.

The last assertion follows from Proposition \ref{lem2} and the fact that, for all $p\in D$ and all sequence $\sigma\subset D$, the following chain of equivalences holds

\[
p\in L_{\tau}(\sigma)\, \iff\, \sigma \hbox{ converges to }p \hbox{ with $\tau$}\, \iff\, \sigma \hbox{ converges to }p \hbox{ with $\tnewd$}\,\iff\,p\in L_{\tnewd}(\sigma),
\]
where the first equivalence follows as $L_\tau$ is the (first order) limit operator associated to $\tau$, the second one from the last assertion of Theorem \ref{prop2} and the third one from the definition of $L_{\tau^*}$. \cvd

\smallskip

\noindent This result provides an elegant solution to our problem.
However, one can wonder for a better understanding of $\tnewdl$. %$(=(\tau^*)_{Seq})$.
Notice that the topology $\tnewdl$ is defined in terms
of the limit operator $L_{\tau^*}$, which is not directly based on
some limit operator $L$ of $\tau$, but on the topology $\tau^*$.
This is a difficulty in order to manage such a topology for practical
purposes.
%In fact, we have not obtained yet that $\tau$ and
%$\tau_*$ must agree on $D$ (remarkably, the analog property was
%proved in Theorem \ref{prop2}). The next subsection will remedy
%this.

\subsection{Limit operators and further properties of the separating topology}\label{sec3.3}

This subsection is devoted to find an alternative limit operator
$L^*$, directly constructed from $L$, whose derived topology
$\tau_{L^*}$ coincides with the {\crustL} one
$\tnewdl$. The natural candidate is:
\begin{equation}\label{def2}
\lnewd (\sigma):=\left\{\begin{array}{lc}L(\sigma)\cap D &
\hbox{if $\exists\, \kappa\subset \sigma$ and $p\in D$ such that }
p\in L(\kappa)  \\ L(\sigma) & \hbox{otherwise}
\end{array}\right.
\end{equation}
(so, $\lnewd (\sigma)=L(\sigma)\cap D$ not only when $L(\sigma)\cap D\neq\emptyset$ but also when $L(\kappa)\cap D\neq\emptyset$ for some subsequence $\kappa$; this is necessary in order to ensure that
$L^*$ is a limit operator, see Proposition \ref{prop1'}). Indeed,
the definition of $L^*$ suggests that it is the smallest
modification of $L$ such that no sequence $\sigma \subset X$ will
converge to both, $p\in D$ and $q\in X\backslash D$. However, a
caution must be taken into account: this  property will be ensured
only if $L^*$ is of first order
---otherwise, $\sigma$ might $\tau_{L^*}$-converge to both $p$ and
$q$, even if $q\not\in L(\sigma)$. In order to ensure that
$L^*$ is of first order, a natural requirement will be to assume
that so is $L$. Of course, this is not restrictive from a
fundamental viewpoint, as one can always replace it by the associated limit
operator $L_{\tau}$ in \eqref{eq11}, which is of first order. However, this
caution must be taken into account from a practical viewpoint,
when $L^*$ is being computed from a sequential topology $\tau$
constructed by means of some prescribed limit operator $L$ (as in the case
of some commented relativistic boundaries).

With this aim for $L^*$, notice first that formula \eqref{*}
in Lemma \ref{lem3.3a} can be regarded as a characterization of
the property (\AFin) of Defn. \ref{dcrust} in terms of the limit
operator $L$. It is convenient to rewrite also the condition
%(A$_1$) and
(\Asep) in such terms as follows.

\begin{lemma}\label{lem3.3b} Let $L, L'$ be two limit operators on $X$ with derived topologies $\tau_L %(\equiv \tau)
$ and $\tau_{\laux}$ resp., and let  $D$ be open, locally compact
and Hausdorff for $\tau_L$. Assume also that $\tau_L \subset
\tau_{\laux}$.

If the topology $\tau_{\laux}$ satisfies the condition
(\Asep) (with $\tnewd\equiv \tau_{\laux}$,
$\tau\equiv \tau_L$), then one has:
\begin{equation}\label{**}
\hbox{when $ \laux(\sigma) \cap D \neq \emptyset$ for some
$\sigma\subset X$, then $\laux(\sigma)\subset D$.}
\end{equation}
The converse is true if we assume that $L$ is of first order on
$D$  and, additionally:
\begin{equation}\label{e5}
\hbox{$\ltiD(\sigma)=\laux\mid_{D}(\sigma)$ for any sequence
$\sigma\subset D$}
\end{equation}
(so that $\tau_L |_D = \tau_{\laux} |_D$ by Proposition
\ref{lem2}).
\end{lemma}
{\Demo} To the right, observe that if \eqref{**} does not hold,
then there exist $p\in D, q\in X\setminus D$ and a sequence
$\sigma\subset X$ such that $p,q\in \laux(\sigma)$. Then, from
\eqref{eq1}, $\sigma$ converges with $\tau_{\laux}$ to both $p,q$,
which contradicts that $\tau_{\laux}$ satisfies
(\Asep).

For the converse, for each $p\in D$, take some compact (for $\tau_L$ of, equally by \eqref{e5}, for $\tau_{L'}$)
neighborhood $K\subset D$ of $p$, and let $U\in \tau_{L}$ be its
$\tau_{L}$-interior.
%Notice that, by the
%Hausdorfness of $D$, $K$ is $\tau_L$-closed in $D$ (so, it
%coincides with the $\tau_L$-closure of $U$ in $D$)\footnote{!!: He
%dejado la frase de arriba hasta ahi' (aunque no creo que sea
%estrictamente necesaria), pero he suprimido: and, thus, there
%exists some $\tau_L$-closed set $C$ such that $K=C\cap D$ [por si
%fuera necesario restaurarla, pongo otra nota a pie abajo]}, and
The property (\Asep) will follow trivially
if we prove that $X\backslash K$ is open for $\tau_{\laux}$, that
is, if $\laux(\sigma)\subset K$ for any sequence $\sigma\subset
K$. So, assume by contradiction that $q\in \laux(\sigma)\setminus
K$ for some sequence $\sigma\subset K$. Then $\sigma\rightarrow q$
with $\tau_{\laux}$ and, therefore, with $\tau_L$. Note that $K$
is a compact subset of $D$ which contains $\sigma$, and both, $D$
and $K$ are sequential spaces with the restriction of $\tau_L$
(use either \cite[Prop. 1.9]{Franklin} or Proposition \ref{lem2}
for the sequentiality of $D$ and, then, \cite[Lemma 3.7]{Goreham}
for $K$). So, the compactness of $K$ implies its sequential
compactness (see the end of Remark \ref{r2.5}) and there exists a
subsequence $\kappa\subset \sigma$ and some $p'\in K\subset  D$
such that $\kappa\rightarrow p'$ with $\tau_{L}$. Since $L$ is of
first order on $D$, necessarily $p'\in L(\kappa)$ which, combined
with (\ref{e5}), implies $p'\in \laux(\kappa)$. On the other hand,
by hypothesis, $q\in \laux(\sigma)\setminus K$. So, there are two
possibilities: either $q\in D$, which contradicts the
Hausdorffness of $\tau_{\laux} |_D=\tau_L |_D$, or $q\in
X\setminus D$, which contradicts property (\ref{**}) applied to
$\kappa$. \cvd

\smallskip

%\begin{remark}{\em
%Even admitting the first order character of $L$ on $D$ as an
%ambient hypothesis, the condition \eqref{**} in Lemma
%\ref{lem3.3b} is not a complete characterization of
% condition (A$_{\hbox{\scriptsize Sep}}$), as this condition  does not imply (\ref{e5})
% (say, $\tau_{\laux}$ might satisfy condition (A$_{\hbox{\scriptsize Sep}}$),
% but still include additional open subsets in $D$). Nevertheless, the condition
% \eqref{**} will be used under  a
%minimality hypothesis for $\tau_{\laux}$ which will imply
%(\ref{e5}), and this will be enough for our purposes.\footnote{J:
%Quizás tendría que hablar con Miguel para aclarar alguna cosa
%aqui. M. La he cambiado un poco para ser ma's directo, pero no
%estoy seguro ahora de si es exactamente cierto lo que ahora digo
%de \eqref{e5} (y que tal vez no me atrevi' a decir antes). Lo
%hablamos y, si es necesario, lo quitamos/modificamos. }
%}\end{remark}

The hypotheses under which the condition \eqref{**} characterizes
(\Asep) are technical, but they will become natural in the applications of Lemma \ref{lem3.3b}.
Next, let us  characterize $\lnewd$ for any limit operator $L$.
%
%With this previous lemmas, we are able to obtain an analogous version of Theorem \ref{prop2} but restricted to sequential spaces.
\begin{proposition}\label{prop1'}
Let $(X,\tau_L)$ be a sequential topological space and $D\subset
X$ a Hausdorff locally compact open subset of $X$. Then, $\lnewd$
is a limit operator which satisfies the properties (\ref{*}),
(\ref{**}) and (\ref{e5}), and it is the maximum operator satisfying them, that is: if $L'$ is another limit operator
satisfying (\ref{*}), (\ref{**}) and (\ref{e5}), then $\laux \subset \lnewd$ (i.e.,
$\laux(\sigma)\subset \lnewd(\sigma)$ for any sequence
$\sigma\subset X$).
\end{proposition}
\noindent {\it Proof.} First, let us prove that $\lnewd$ is a
limit operator, i.e. it satisfies (\ref{***}). Assume by
contradiction that there exist a sequence $\sigma$ and a
subsequence $\kappa$ such that $\lnewd(\sigma)\not\subset
\lnewd(\kappa)$. Recall that $L^{*}(\sigma)\subset
L(\sigma)\subset L(\kappa)$. Hence, necessarily $L^{*}(\kappa)\neq
L(\kappa)$. In particular, $L(\kappa)\cap D\neq \emptyset$, and
thus, $L^{*}(\sigma)=L(\sigma)\cap D\subset L(\kappa)\cap
D=L^{*}(\kappa)$, in contradiction with the initial hypothesis.
%Taking into account that
%$L(\sigma)\subset L(\kappa)$ (recall that $L$ was a limit
%operator), necessarily necessarily $L^*(\kappa)\varsubsetneq
%L(\kappa)$ and $L^*(\kappa)= L(\kappa)\cap D $. But then
%$L^*(\sigma)= L(\sigma)\cap D $ and so $L^*(\sigma)\subseteq
%L^*(\kappa)$. From the latter, $L(\kappa)\cap D\neq \emptyset$,
%and thus, $L^*(\sigma)\neq L(\sigma)$, which is a contradiction
%with the former.

%and the definition of $\lnewd$, a discussion of cases yields
%$\lnewd(\sigma)=L(\sigma)\varsupsetneq L(\sigma)\cap D$. But,
%then, $\sigma$ (and, therefore, $\kappa$) cannot admit a
%subsequence $\kappa'$ with some $p\in L(\kappa')\cap D$.
%Therefore, we also have $\lnewd(\kappa)=L(\kappa)$, a
%contradiction.

From the definition of $L^*$, it satisfies the properties
\eqref{*}, \eqref{**} and \eqref{e5}. So, it remains to prove the
maximal character of $L^*$. Let $\laux$ be another limit operator
satisfying such properties. In order to prove that
$\laux(\sigma)\subset\lnewd(\sigma)$ for any sequence
$\sigma\subset D$, consider the cases:

\begin{itemize}
\item There exists $\kappa\subset \sigma$ and $p\in D$ such that
$p\in L(\kappa)$. Observe that, as $D$ is open, we can assume
$\kappa\subset D$. Then, $\lnewd(\sigma)=L(\sigma)\cap D$ from the
definition \eqref{def2} and $p\in\laux(\kappa)$ as $\laux$
satisfies (\ref{e5}). Assume by contradiction that
$\laux(\sigma)\not\subset \lnewd(\sigma)$. There exists $q\in
\laux(\sigma)$ such that $q\not\in \lnewd(\sigma)$ and,
necessarily $q\not\in D$ by (\ref{e5}).
%\footnote{!! En lugar de
%by (\ref{e5}) ponia: (otherwise $q\in \laux(\sigma)\subset
%L(\sigma)$ implies $q\in L(\sigma)\cap D=\lnewd(\sigma)$, a
%contradiction)}.
So, $q$ belongs to both $\laux(\kappa)$ and
$X\setminus D$, which absurd as $p\in \laux(\kappa)\cap D$ and
$\laux$ satisfies (\ref{**}).

\item There are no $\kappa\subset \sigma$
under previous conditions. Then, $\lnewd(\sigma)=L(\sigma)$, and,
from (\ref{*}), one has $\laux(\sigma)\subset
L(\sigma)=\lnewd(\sigma)$. \cvd
\end{itemize}
The following technical assertion will be useful; notice that Lemma \ref{lem3.3b} gives conditions for its applicability.
\begin{lemma}\label{ll} Assume that $\tau_{\lnewd}$ satisfies
%\footnote{M: He agnadido ``$(A_1)$ and'' en consonancia con lo que dije el el Lemma \ref{lem3.3}.
%No he hecho ma's modificacio'n aqui'.J: Yo diría que no hace falta $(A_1)$ aqui...(Jose Luis opina lo mismo).} $(A_1)$ and
(\Asep). If $\sigma\rightarrow q$ with
$\tau_{\lnewd}$ and $q\in L(\sigma)$, then $q\in \lnewd (\sigma)$.
\end{lemma}

\Demo If $q\in D$ then $q\in L(\sigma)\cap D=L^*(\sigma)$, as
required. So, assume that $q\in X\setminus D$. It suffices to
prove that $L(\kappa)\cap D=\emptyset$ for any subsequence
$\kappa\subset\sigma$ (since then $q\in L(\sigma)=L^*(\sigma)$).
Assuming by contradiction that there exists $p\in L(\kappa)\cap
D\neq\emptyset$, the sequence $\kappa$ converges to both, $p$ and
$q$, with $\tau_{L^*}$, in contradiction with property
(\Asep). \cvd

\smallskip

Now, we can establish the following result that clarifies the role of limit operators:
\begin{theorem}\label{prop1}
Let $(X,\tau\equiv \tau_L)$ be a sequential topological space and
let $D\subset X$ be a Hausdorff locally compact open subset of
$X$. Assume that the limit operator $L$ is of first order on $D$. Then, the following statements hold:

\begin{itemize}
\item[(i)] The topology derived by the limit operator $\lnewd$ defined on (\ref{def2}) is finer than the unique sequentially minimally D-separating topology. So, we have the following chain of topologies:
\begin{equation}\label{cadenatop}
\tau\subset \tau^*\subset\tnewdl\subset \tau_{\lnewd}.
\end{equation}
\item[(ii)] If $L$ is of first order (on all $X$), then $\lnewd$ is also of first order and both topologies $\tnewdl$ and $\tau_{\lnewd}$ coincide.

\end{itemize}
\end{theorem}

\noindent \Demo For (i), let us consider a set $C$ which is closed set for the topology $\tnewdl$ and prove that such a set is also closed for $\tau_{\lnewd}$. Our aim is to show that $\lnewd(\sigma)\subset C$ for all sequence $\sigma\subset C$. Consider a point $p\in \lnewd(\sigma)$ and recall that, from the definition of $\lnewd$, we may consider two cases. For the first one, assume that $\lnewd(\kappa)\cap D\neq \emptyset$ for some subsequence $\kappa\subset \sigma$, and so, that $p\in \lnewd(\sigma)=L(\sigma)\cap D$. Then, $\sigma$ converges to $p$ with $\tau$ and $\sigma$ will abandon any compact set of $D$ not containing the point $p$ (recall that $D$ is Hausdorff). Therefore, $\sigma$ converges to $p$ with $\tau^*$ and, from the definition of its limit operator, $p\in L_{\tnewd}(\sigma)$. But $\tnewdl=(\tau^*)_{Seq}$ (Theorem \ref{t1}) and $C$ is closed for $\tnewdl$, so $p\in C$.

For the second case, assume that $L(\kappa)\cap D=\emptyset$ for all subsequence $\kappa\subset\sigma$, and so, $p\in X\setminus D$. As $p\in \lnewd(\sigma)$, we also have that $p\in L(\sigma)$, and so, $\sigma$ converges to $p$ with $\tau$. Moreover, such a sequence will converge to $p$ also with $\tnewd$: otherwise, there would be a compact $K$ and a subsequence $\kappa\subset K$ of $\sigma$ converging to a point $q\in D$ with the topology $\tau$. However, as $L$ is of first order on $D$, this would imply that $q\in L(\kappa)\cap D\neq\emptyset$, a contradiction. So, $\sigma$ converges to $p$ with $\tnewd$ and, thus, $p\in L_{\tnewd}(\sigma)$. In conclusion, as $C$ is closed for $\tnewdl$, again $p\in C$, as desired. To finish (i), recall that the two first inclusions in \eqref{cadenatop} are obvious from (\AFin) in Defn. \ref{dcrust} and Theorem \ref{t1} (notice also Prop. \ref{p} (b)), respectively.

\smallskip

Now, let us prove (ii). For the first order character of $\lnewd$, observe first that since $L^*$ satisfies properties (\ref{*}),
(\ref{**}) and (\ref{e5}) (recall Proposition \ref{prop1'}), and
$L$ is of first order, Lemmas \ref{lem3.3a}, \ref{lem3.3b}, ensure
that $\tau_{L^*}$ satisfies (\AFin) and (\Asep). Now assume that
$\sigma\rightarrow q$ with $\tau_{L^*}$ for some sequence
$\sigma\subset X$. From (\AFin), $\sigma\rightarrow q$ with
$\tau_L$ and, as $L$ is of first order, $q\in L(\sigma)$. Then,
Lemma \ref{ll} applies, and $q\in \lnewd(\sigma)$ follows, as
desired.

For the second assertion of (ii),  recall that from (i), we have the inclusion $\tnewdl\subset \tau_{\lnewd}$. As $L_{\tnewd}$ is a first order limit operator, Lemma \ref{lem3.3a} ensures that $\lnewd(\sigma)\subset L_{\tnewd}(\sigma)$ for all sequence $\sigma\subset X$. So, taking into account Proposition \ref{prop1'}, it suffices to prove that
$L_{\tau^*}$ satisfies properties (\ref{*}), (\ref{**}) and
(\ref{e5}).  Properties (\ref{*})
and (\ref{**}) follow from properties (\AFin) and
(\Asep) (recall Lemmas \ref{lem3.3a},
\ref{lem3.3b}) while (\ref{e5}) is a consequence of Theorem \ref{t1}. \cvd

\begin{remark}\label{r} {\em
As announced at the beginning of the present subsection, the
hypotheses of being $L$ of first order is necessary in Theorem
\ref{prop1} (even though it is not in Proposition \ref{prop1'}),
but this is not a fundamental restriction because, given a
sequential topology $\tau$, one can consider its (first order)
limit operator $L_\tau$ given in \eqref{eq11}.
%So, Theorem
%\ref{prop1} becomes in particular the expected improvement of
%Theorem \ref{t1}:

%\begin{quote}
% For $(X,\tau)$ sequential  and any open $D\subset X$  as
% above,
% the unique  {\crustL} topology $\tau_*=\tau_{L_{\tau^*}}$ satisfies that   the restrictions of $\tau$ and
%$\tau_{L_{\tau^*}}$ on $D$ coincide. \end{quote}

Nevertheless, Theorem \ref{prop1} allows to understand better the role of $L$ and $L^*$ in order to obtain the separating topologies. Summing up, associated to a sequential topology $\tau$ writtten as
$\tau=\tau_{L}$ we have defined three  different topologies:
$\tnewd$, $\tnewdl(=(\tau^*)_{Seq})$ and $\tau_{\lnewd}$. They always
satisfy
$$
\tau \subset \tnewd \subset \tnewdl
$$
and, whenever $\tau_{\lnewd}$ satisfies (\Asep) (in particular, when $L$ is of first order on $D$), $
\tnewdl\subset \tau_{\lnewd}$. The first constructed topology $\tnewd$
satisfies the required separating properties, but it is not
necessarily sequential. The second $\tnewdl$ satisfies both, the separating properties and
sequentiality, even though it has a
drawback from the practical viewpoint, namely, one does not have a
priori an explicit expression for $L_{\tnewd}$ in terms of $L$.
Finally, the third $\tau_{L^*}$ may depend on the choice of $L$
for the topology $\tau$, but it coincides with
$\tnewdl$ when $L$ is of first order (i.e. $L=L_\tau$). So, one obtains the explicit
limit operator $L_{\tnewd}=L^*$.}

\end{remark}

\begin{figure}
\centering \ifpdf
  \setlength{\unitlength}{1bp}%
  \begin{picture}(435.73, 206.95)(0,0)
  \put(0,0){\includegraphics{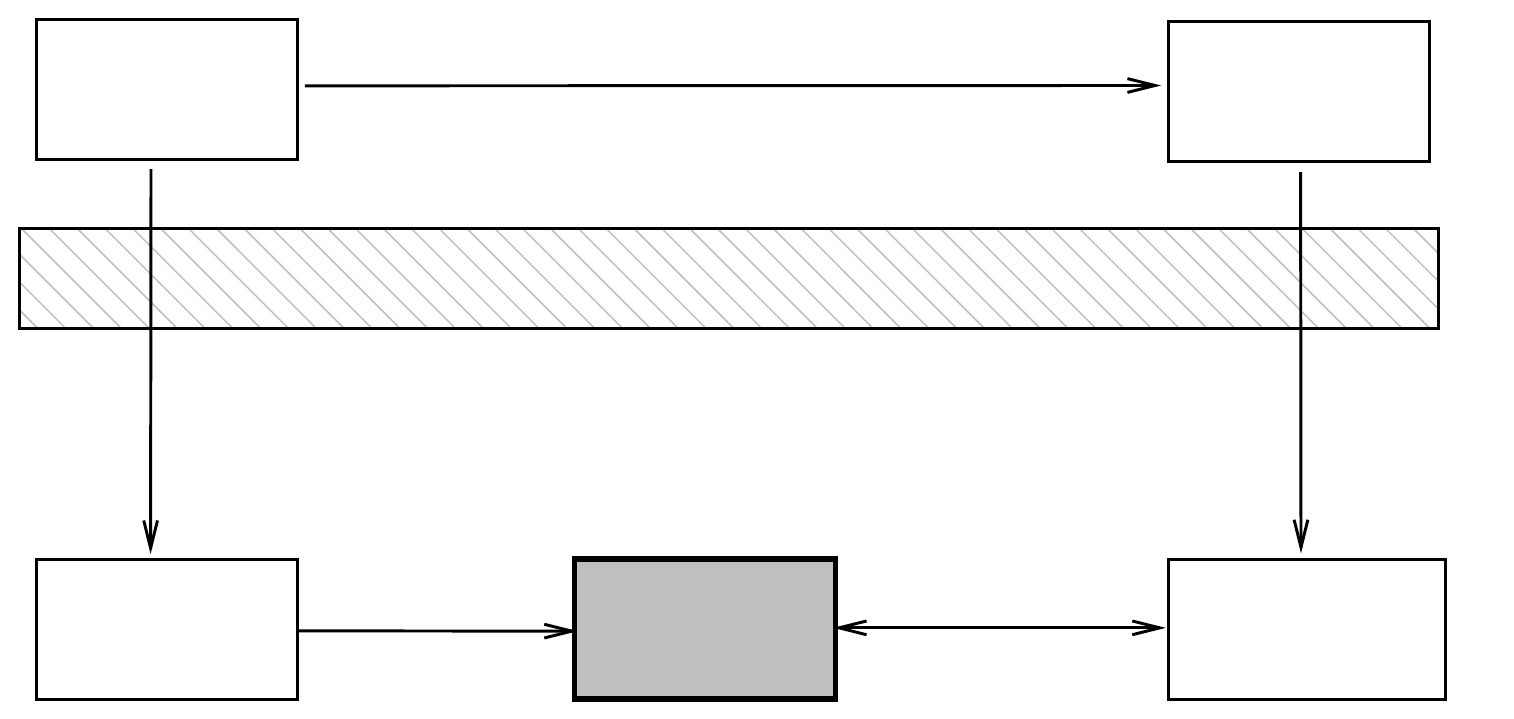}}
  \put(15.43,184.38){\rotatebox{0.00}{\fontsize{7.05}{8.47}\selectfont \smash{\makebox[0pt][l]{General topology}}}}
  \put(155.58,123.12){\fontsize{11.76}{14.11}\selectfont CHOICE OF $D\subset X$}
  \put(11.63,29.66){\fontsize{7.05}{8.47}\selectfont Minim. D-separating}
  \put(45.45,82.89){\fontsize{7.11}{8.54}\selectfont Theorem \ref{prop2}}
  \put(29.62,19.48){\fontsize{7.05}{8.47}\selectfont  topology $\tnewd$}
  \put(169.13,33.43){\fontsize{7.05}{8.47}\selectfont Minim. sequentially}
  \put(179.04,24.41){\fontsize{7.05}{8.47}\selectfont D-separating }
  \put(171.72,15.25){\fontsize{7.05}{8.47}\selectfont topol. $\tnewdl=\tau_{L_{\tnewd}}$}
  \put(356.13,36.58){\fontsize{7.05}{8.47}\selectfont $\tau_{\lnewd}$ with $\lnewd$}
  \put(350.84,28.16){\fontsize{7.05}{8.47}\selectfont defined from $L$}
  \put(338.19,18.19){\fontsize{7.05}{8.47}\selectfont (Maximal limit oper.}
  \put(178.73,190.39){\fontsize{9.96}{11.95}\selectfont when $\tau$ is sequential}
  \put(347.95,183.39){\fontsize{7.05}{8.47}\selectfont $\tau=\tau_{L}$ for some}
  \put(347.63,173.12){\fontsize{7.05}{8.47}\selectfont  limit operator $L$}
  \put(247.19,18.23){\fontsize{6.35}{7.62}\selectfont Coincide if $L$ of first order}
  \put(247.52,10.65){\fontsize{6.35}{7.62}\selectfont ($L=L_\tau$), Theorem \ref{prop1}}
  \put(378.40,79.94){\fontsize{7.11}{8.54}\selectfont Defn. in}
  \put(378.40,70.94){\fontsize{7.11}{8.54}\selectfont form. (\ref{def2})}
  \put(102.50,16.47){\fontsize{6.35}{7.62}\selectfont Theorem \ref{t1}}
  \put(44.15,174.16){\fontsize{7.05}{8.47}\selectfont  $\tau$}
  \put(340.60,11.72){\fontsize{7.05}{8.47}\selectfont --sense of Prop. \ref{prop1'})}
  \end{picture}%
\else
  \setlength{\unitlength}{1bp}%
  \begin{picture}(435.73, 206.95)(0,0)
  \put(0,0){\includegraphics{esquema1.pdf}}
   \put(15.43,184.38){\rotatebox{0.00}{\fontsize{7.05}{8.47}\selectfont \smash{\makebox[0pt][l]{General topology}}}}
   \put(155.58,123.12){\fontsize{11.76}{14.11}\selectfont CHOICE OF $D\subset X$}
   \put(11.63,29.66){\fontsize{7.05}{8.47}\selectfont Minim. D-separating}
   \put(45.45,82.89){\fontsize{7.11}{8.54}\selectfont Theorem \ref{prop2}}
   \put(29.62,19.48){\fontsize{7.05}{8.47}\selectfont  topology $\tnewd$}
   \put(169.13,33.43){\fontsize{7.05}{8.47}\selectfont Minim. sequentially}
   \put(179.04,24.41){\fontsize{7.05}{8.47}\selectfont D-separating }
   \put(171.72,15.25){\fontsize{7.05}{8.47}\selectfont topol. $\tnewdl=\tau_{L_{\tnewd}}$}
   \put(356.13,36.58){\fontsize{7.05}{8.47}\selectfont $\tau_{\lnewd}$ with $\lnewd$}
   \put(350.84,28.16){\fontsize{7.05}{8.47}\selectfont defined from $L$}
   \put(338.19,18.19){\fontsize{7.05}{8.47}\selectfont (Maximal limit oper.}
   \put(178.73,190.39){\fontsize{9.96}{11.95}\selectfont when $\tau$ is sequential}
   \put(347.95,183.39){\fontsize{7.05}{8.47}\selectfont $\tau=\tau_{L}$ for some}
   \put(347.63,173.12){\fontsize{7.05}{8.47}\selectfont  limit operator $L$}
   \put(247.19,18.23){\fontsize{6.35}{7.62}\selectfont Coincide if $L$ of first order}
   \put(247.52,10.65){\fontsize{6.35}{7.62}\selectfont ($L=L_\tau$), Theorem \ref{prop1}}
   \put(378.40,79.94){\fontsize{7.11}{8.54}\selectfont Defn. in}
   \put(378.40,70.94){\fontsize{7.11}{8.54}\selectfont form. (\ref{def2})}
   \put(102.50,16.47){\fontsize{6.35}{7.62}\selectfont Theorem \ref{t1}}
   \put(44.15,174.16){\fontsize{7.05}{8.47}\selectfont  $\tau$}
   \put(340.60,11.72){\fontsize{7.05}{8.47}\selectfont --sense of Prop. \ref{prop1'})}
  \end{picture}%
\fi\caption{\label{figresumen}Summary of the constructions and
results of Section \ref{s}.}
\end{figure}

Although the first order restriction for $L$ is not fundamental in this context (see the previous remark), we will consider also  non-necessarily first order limit operator $L$, as this might be the case for the natural operator for the chronological topology. %(see Section \ref{s5})
So, the next proposition gives a small extension of the previous theorem, to be used later. 
 %More precisely, we will be interested in understanding at what extent the topologies $\tnewdl$ and $\tau_{\lnewd}$ differ under the weaker hypothesis that only the operator $L^*$ (constructed directly from $L$) is of first order. As we will see, even if it is conceivable that such topologies may be different, the cases where this occurs are very artificial. In fact, such a property is only possible when $L$ is not of $k$-th order for any ordinal $k$.

\smallskip

In what follows, $\tau$ will be a sequential topology derived from some limit operator $L$, i.e. $\tau=\tau_L$.

\begin{lemma}\label{previo}
 If $p\in X\setminus D$ satisfies $p\in L^*_{\tau}(\sigma)$ for some sequence $\sigma\subset X$, then $L^i(\sigma)=(L^*)^i(\sigma)$ for any ordinal $i$.
\end{lemma}
{\it Proof:} The inclusion to the left is trivial, so we will focus on the inclusion to the right. We will proceed by transfinite induction. For $i=1$, the equality follows from the definition of $L^*$ and the fact that no subsequence $\kappa$ of $\sigma$ satisfies $d\in L(\kappa)\cap D$; in fact, otherwise, $d\in L_{\tau}(\kappa)$, and thus, $L_{\tau}^*(\sigma)\subset D$, which implies $p\not\in L^*_{\tau}(\sigma)$, a contradiction. So, let us assume that the result is valid for any $j<i$, and let us prove it for $i$. Recall first that $L(\kappa)\cap D=\emptyset$ for any subsequence $\kappa$ of $\{p_n\}\subset \cup_{j<i}L^{j}(\sigma)$. In fact, otherwise, there exists some $d\in L(\kappa)\cap D$ for some subsequence $\kappa$ as before, and thus $d\in L^i(\sigma)\subset L_{\tau}(\sigma)$ (recall Proposition \ref{aux1}), which implies $L_{\tau}^{*}(\sigma)\subset D$, in contradiction to the fact that $p\in L^*_{\tau}(\sigma)$. So, consider a point $q\in L^{i}(\sigma)$. Then, there exists a sequence $\{p_n\}\subset \cup_{j<i}L^{j}(\sigma)=\cup_{j<i}(L^*)^{j}(\sigma)$ (recall the hypothesis of induction) such that $q\in L(\{p_n\})$.
Since $L(\kappa)\cap D=\emptyset$ for any subsequence $\kappa$ of $\{p_n\}$, necessarily $q\in L(\{p_n\})=L^*(\{p_n\})$, and thus, $q\in (L^*)^{i}(\sigma)$. \cvd
\smallskip

%\begin{remark} Notice that $(L^*)^i(\sigma)= (L^i)^*(\sigma)$ always. In fat, it is enough to
%$ L^i(\kappa)\cap D=\emptyset$ para toda subsucesión
%\kappa\subset\sigma. Supongamos por reducción al abusurdo que d\in
%L^i(\kappa)\cap D para alguna subsucesión \kappa\subset\sigma. Entonces
%d\in L^i(\kappa)\subset L_{\tau}(\kappa) (Prop. 2.8). Esto contradice que
%p\not\in D y p\in L_{\tau}^*(\sigma)
%\end{remark}

\begin{proposition}\label{prop4.2}
 Let $L$ be a limit operator of $k$-th order (for some ordinal $k$) which is also of first order on $D$. If $L^*$ is of first order then $L^*=L^*_{\tau}$ (where $L^*_{\tau}:=(L_{\tau})^*$).
\end{proposition}
{\it Proof:} Since $L(\sigma)\subset L_{\tau}(\sigma)$ for all $\sigma\subset X$, and $L$ is of first order on $D$, we directly have
\[
L^*(\sigma)\subset L^*_{\tau}(\sigma).
\]
For the inclusion to the left, consider a point $p\in L^*_{\tau}(\sigma)$. If $p\in D$ then
\[
p\in L^*_{\tau}(\sigma)\;\; \Rightarrow\;\; p\in L_{\tau}(\sigma)\;\; \Rightarrow\;\; p\in L(\sigma)\;\; \Rightarrow\;\; p\in L^*(\sigma),
%\quad\hbox{as required}
\]
as required (we have used (\ref{def2}) for the first and third implications, and the first order character of $L$ on the open set $D$ for the second implication). If, otherwise $p\in X\setminus D$, from Lemma \ref{previo}, $L^i(\sigma)=(L^*)^{i}(\sigma)$ for any ordinal $i$. Moreover, as $L^*$ is of first order, $L^*(\sigma)=(L^*)^{i}(\sigma)$ for any ordinal $i$ (recall Corollary \ref{cc}). Therefore, taking into account that $L$ is a limit operator of $k$-th order, we have
\[
p\in L_{\tau}^*(\sigma) \subset L_{\tau}(\sigma)=L^k(\sigma)=(L^*)^{k}(\sigma)=L^*(\sigma).
\]
In conclusion, in both cases we deduce that $p\in L^*(\sigma)$, as desired. \cvd\\

\section{Completions of spacetimes, $T_2$-separability and c-boundary} \label{s4}

\subsection{A revision of some boundaries in  Relativity} \label{s4.1}
Let $(M,g)$ be a spacetime, i.e. a connected time-oriented
Lorentzian manifold. There are several types of boundaries  in
Relativity applicable to (some classes of) spacetimes, yielding a
completion  of the spacetime $\overline{M}=M\cup \partial M$.
Among them, one has Geroch {\em geodesic boundary} (g-boundary)
\cite{GeJMP}, Schmidt {\em bundle boundary} (b-boundary) \cite{Sc,
Sc2}, Scott and Szekeres {\em abstract boundary} (a-boundary)
\cite{SS}, Penrose {\em conformal boundary} \cite{HE, Wa} and
Geroch, Kronheimer and Penrose {\em causal  boundary}
(c-boundary).

The g-boundary was defined by using classes of incomplete
geodesics in $(M,g)$. For  the b-boundary, one defines a certain
positive definite metric on the bundle of linear frames $LM$  of
$M$, takes the Cauchy completion of $LM$ and induces then a
boundary for $(M, g)$.  Both constructions satisfy the following a
priori desirable properties pointed out by Geroch, Liang and Wald
\cite{GLW}:

(i) every incomplete geodesic in the original spacetime terminates
at a point, and

(ii) they are geodesically continuous, in a sense rigorously
defined in \cite{GLW}.

%\smallskip

\begin{figure}
\centering
\ifpdf
  \setlength{\unitlength}{1bp}%
  \begin{picture}(154.66, 240.57)(0,0)
  \put(0,0){\includegraphics{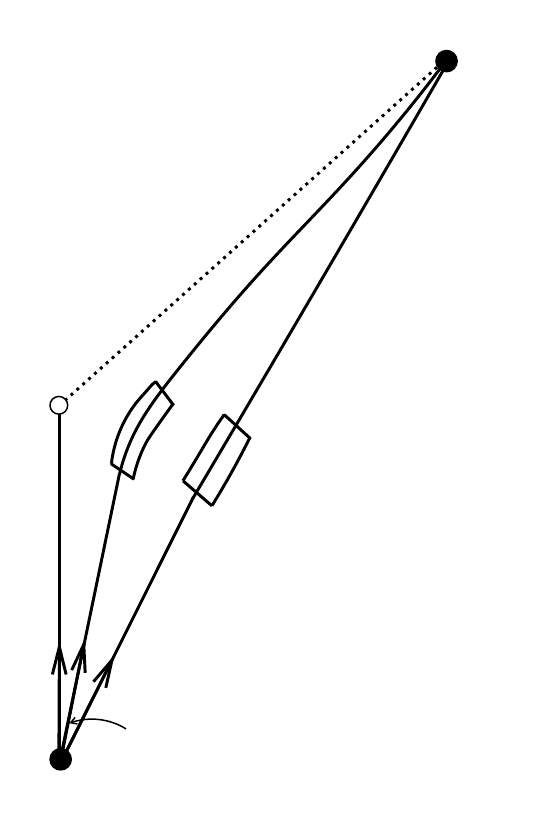}}
  \put(87.69,140.79){\fontsize{8.54}{10.24}\selectfont $\gamma_1$}
  \put(66.86,148.17){\fontsize{8.54}{10.24}\selectfont $\gamma_2$}
  \put(7.83,86.99){\fontsize{8.54}{10.24}\selectfont $\gamma$}
  \put(68.28,100.34){\fontsize{8.54}{10.24}\selectfont $U_1$}
  \put(44.51,111.68){\fontsize{8.54}{10.24}\selectfont $U_2$}
  \put(14.62,8.35){\fontsize{11.38}{13.66}\selectfont $p$}
  \put(11.96,127.34){\fontsize{11.38}{13.66}\selectfont $s$}
  \put(129.91,226.24){\fontsize{11.38}{13.66}\selectfont $r$}
  \put(31.70,41.16){\fontsize{8.54}{10.24}\selectfont $\xi_1$}
  \put(35.85,26.20){\fontsize{8.54}{10.24}\selectfont $\xi_2$}
  \put(5.67,35.22){\fontsize{8.54}{10.24}\selectfont $\xi$}
  \end{picture}%
\else
  \setlength{\unitlength}{1bp}%
  \begin{picture}(154.66, 240.57)(0,0)
  \put(0,0){\includegraphics{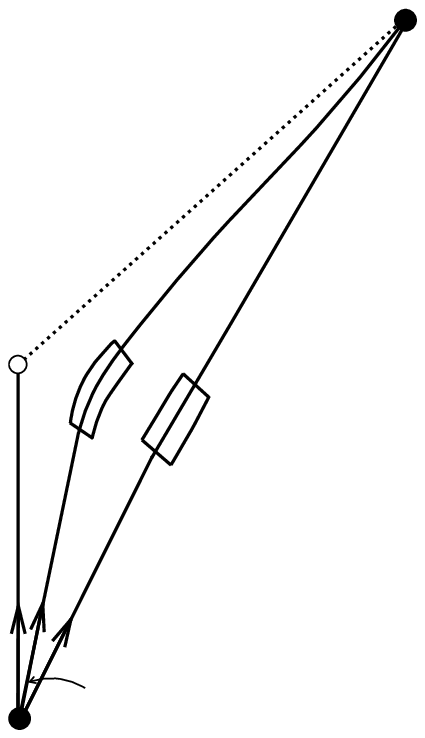}}
  \put(87.69,140.79){\fontsize{8.54}{10.24}\selectfont $\gamma_1$}
  \put(66.86,148.17){\fontsize{8.54}{10.24}\selectfont $\gamma_2$}
  \put(7.83,86.99){\fontsize{8.54}{10.24}\selectfont $\gamma$}
  \put(68.28,100.34){\fontsize{8.54}{10.24}\selectfont $U_1$}
  \put(44.51,111.68){\fontsize{8.54}{10.24}\selectfont $U_2$}
  \put(14.62,8.35){\fontsize{11.38}{13.66}\selectfont $p$}
  \put(11.96,127.34){\fontsize{11.38}{13.66}\selectfont $s$}
  \put(129.91,226.24){\fontsize{11.38}{13.66}\selectfont $r$}
  \put(31.70,41.16){\fontsize{8.54}{10.24}\selectfont $\xi_1$}
  \put(35.85,26.20){\fontsize{8.54}{10.24}\selectfont $\xi_2$}
  \put(5.67,35.22){\fontsize{8.54}{10.24}\selectfont $\xi$}
  \end{picture}%
\fi
\caption{\label{figGLW} In their  example, Geroch, Liang and Wald considered $\R^2\setminus \{s\}$ endowed with a metric $\Omega\cdot \eta$, where $\eta$ is the Minkowski metric and $\Omega$ is a conformal factor. The conformal factor was defined in such a way that all $\gamma_i$ are geodesics and $\Omega\equiv 1$ outside of the open sets $U_i$ as well as on each
curve $\gamma_i$ inside $U_i$ (see \cite{GLW} for details).
}
\end{figure}

\noindent However, Geroch. et al. found
 a  drawback  for any boundary satisfying (i) and (ii): in a simple insightful example of
(stably causal) spacetime, these two conditions implied   the
existence of a point $r$ in the spacetime and a point $s$ in the boundary
non-$T_1$-separated (see Fig. \ref{figGLW});
the authors proposed even a refined version
of the example that was flat. The topology in the examples was
always sequential and, moreover, there was  a natural intuitive
sense of convergence in the formulation of (ii). So, this drawback
was regarded as a reason to reject a priori the topologies
satisfying them.

Now, recall that our previous results can be applied. More precisely, putting $X$ equal to the union of
$M$ and its g- or b- (or other) boundary $\partial M$, and  $D=M$, one can modify slightly the original topology $\tau$ by taking the \crust one $\tau^*$ provided by Theorem \ref{prop2} (or the {\crustL}
one $\tnewdl$ in Theorem \ref{t1}, if sequentially is also required to be preserved).
% For this purpose, one should check first
%that $M$ is still an open subset in the completion, but even if this did not happen one can %make just a small previous modification: simply,  add  $M$ as an open
%subset in order to generate  the topology of the completion (this would preserve the properties (i) and (ii) above\footnote{?? M. Podriamos agnadir: as well as sequentiality, pues creo que es cierto, pero si hay algu'n problema, no lo agnadimos  Check!}).
%Then, putting $X$ equal to the union of
%$M$ and its g- or b- (or other) boundary $\partial M$, and  $D=M$, one has just to take the topology $\tau^*$ (or $\tau_*$, if sequentially is also required to be preserved).
This topology does not present the commented
drawback and can be developed further. Of course, this
modification of the topology will not solve magically all the
problems of these boundaries, but it suggests possible
improvements and opens the opportunity to re-study them (see, for instance, \cite{MV}).

Let us explain this briefly for the g-boundary (even though the
b-boundary is more elegant and appealing mathematically
\cite{AmGu, Gu},  it has other type of drawbacks from the physical
viewpoint, see \cite{Bo, Jo}). As a previous question, one has to ensure that, if the spacetime is dense for the topology $\tau$ of the original completion, it is also dense for the modified one $\tau^*$. For this purpose, the following straightforward proposition is clearly applicable to the g-boundary, as well as to the
b-, c- and conformal boundaries:

\begin{proposition} \label{plast}
Let $(X,\tau)$ be a %sequential
topological space and $D\subset
X$ a Hausdorff locally compact open subset of $X$. Assume that for
any $x\in X\setminus D$ there exists some sequence $\sigma\subset
D$ such that $x\in L_\tau(\sigma)$. Then, $D$ is
dense in $X$ for $\tau$. If, in addition,  such a $\sigma$ can be chosen such that  it satisfies $L_\tau(\sigma)=L_\tau^*(\sigma)$, then $D$ is also dense for $\tnewdl$.
\end{proposition}
%\Demo The property is obvious for $\tau_L$ because $\sigma$ will
%$\tau_L$-converge to $x$. Moreover, by hypothesis \footnote{Pero,
%si existiese una subsucesion $\kappa$ de $\sigma$ tal que
%$L(\kappa)\cap D\neq\emptyset$ entonces $L^*(\sigma)\neq
%L(\sigma)$?} $L(\sigma)\subset X\setminus D$, hence
%$L^*(\sigma)=L(\sigma)$ and $\sigma$ will $\tau_{L^*}$-converge to
%$x$ too.

%Assume by contradiction the existence of some $U$ in $\tau_L$
%(resp. $\tau_{L^*}$) such that $x\in U\subset X\setminus D$. Let
%$\sigma\subset D$ be such that $x\in L(\sigma)\subset X\setminus
%D$. Then, $L^*(\sigma)=L(\sigma)\subset X\setminus D$, and thus,
%$X\setminus D$ is not closed neither for $\tau_L$ nor for
%$\tau_{L^*}$.

\smallskip

Now, our modified topology $\tau^*$ for
the g-boundary must violate one of the two properties (i) or (ii)
above. Clearly, (i) will not be violated because of the minimality of
our modification (recall the previous proposition).
%Jony: Esto debiera ser general para cualquier completación satisfaciendo (i) y (ii). Por (i) todo punto $p$ del borde se puede expresar como $p=exp_{q}(\xi)$ para un cierto vector $\xi$ y un cierto punto $q$. Sea $\gamma:[0,1)\rightarrow M$ (es abierto en 1 porque el punto está en el borde) la geodésica tal que $\gamma(0)=q$ y $\dot{\gamma}(0)=\xi$ y tomemos una sucesión $t_n\nearrow 1$. Denominemos $\xi_n$ a los vectores tales que $exp_{q}(\xi_n)=\gamma(t_n)$. Es claro que $\xi_n\rightarrow \xi$ por lo que, dado que (ii) pide que la exponencial sea continua en la completación, $\gamma(t_n)\rightarrow p$.}
So, the key is the meaning of the hypothesis {\em geodesically
continuous} in (ii). Geroch, Liang and Wald introduced a natural
definition of this concept in terms of the exponential map. In addition, they
gave a clever counterexample $(M,g)$ to the property of
$T_1$-separability, by exploiting the following
property (see Figure \ref{figGLW} as well as \cite{GLW}): %\footnote{!! Pondria ... property (see Figure \ref{??} as
%well as \cite{GLW}) Y agnadiria la figura 2 de ese articulo}
removing a point $s$ in Lorentz-Minkowski $\LL^2$, they found a
metric $g$ (conformal to the usual one) and two points $p, r\in M$
joined by a sequence of (future-directed) timelike geodesics
$\gamma_i$ of length equal to 1 such that; (a) their initial
velocities $\xi_i=\gamma'_i(0)$ converge to a timelike vector
$\xi$,  and (b) the geodesic $\gamma$ with $\xi=\gamma'(0)$
satisfy that $\lim_{t\nearrow 1}\gamma(t)$ is the removed
point $s$. Their conclusion was that $s$ should be identified with
a point of the g-boundary, and any neighborhood of this point contains $r$. Our modified topology $\tau^*$ separates $r$
and $s$, and seems to give a reasonble behavior for the topology
in this particular example. Nevertheless, this does not simply mean that $\tau^*$ must be the right topology for the g-construction
(notice that Geroch had also suggested in his original article the
possibility of a modification of the g-topology,  see the
footnotes 10 and 14 in \cite{GeJMP}). A closer look to Geroch et
al.'s counterexample shows that they took advantage of the lack of
compactness of $J^+(p)\cap J^-(r)$ as well as the lack of good
convergence properties in the closure of a convex set.  So,  a first
test for a redefinition might be carried out by  restricting the
class of spacetimes (for example, starting at globally hyperbolic
ones).
%Notice that in this spacetimes the Lorentzian distance
% and, eventually, formulating the topology in terms of the
%globally well-defined  (instead of merely
%convex neighborhoods).
We leave this question as open for possible
future studies.

The a-boundary, conformal boundary and c-boundary are not affected
a priori by previous objection, as they are not formulated in
terms of geodesics. The a-boundary \cite{SS} is defined in terms of open
embeddings and has been developed
systematically at the level of the set of ideal points.
At the topological level, Barry and Scott have introduced recently two  topologies for the a-boundary, the so-called {\em attached} and {\em strongly attached point topologies} (see \cite{BarryScott2,BarryScott} for details). By construction, these topologies are not affected by the problem of lack of Hausdorffness between manifold and boundary points, and they have other desirable properties  (in fact, their second proposal improves some properties of the first one).
% even though there has been some recent progress and proposals for its topology \cite{BarryScott}.\footnote{?? M. {\bf Por favor, echadle un vistazo a  arXiv:1401.1287 , arXiv:1201.6414 ,  y ESPECIALMENTE A BARRY Y SCOTT CQG'11 (no fuesteis vosotros referees the eso??)}}; our results could be eventually fruitful for proposals in this sense.
%% According to
%our results, if possible attempts to topologize it were fruitful,
%the lack of $T_1$-separability between points of the boundary and
%the spacetime would not be an (a priori) reason for its rejection.
The conformal boundary is defined in terms of open conformal
embeddings in a bigger (and Hausdorff) spacetime; so, it is not
affected by these problems of separability. Nevertheless, harder
problems appear in order to ensure the existence of such conformal
embeddings and, in this case, the uniqueness of the so-obtained
conformal boundary. The c-boundary appears as a natural
alternative to the conformal boundary. In fact, the c-completion
of spacetimes is a conformally invariant and systematic
construction, applicable to any strongly causal spacetime.

\subsection{Basics on the c-boundary} \label{s4.2}

The {\em c-completion} of a strongly causal spacetime is
constructed by adding {\em ideal points} to the spacetime in such
a way that any timelike curve in the original spacetime acquires
some endpoint in the new space. In the original article by Geroch,
Kronheimer and Penrose  \cite{GKP}, some doubts on its topology
were pointed out. In fact, the original definition has suffered
quite a few modifications. We will consider here  the recent
redefinition in \cite{FHS} (which includes an approach to the
topology of partial boundaries by Harris \cite{H1, H2}), and refer
also there for extensive bibliography on the topic (see also
\cite{H3,FHSisocausal,FHSst} and references therein). This
c-completion of a spacetime is always $T_1$, and there are
situations where the possible non-$T_2$ separation of two points of the boundary
appears as natural. However, there exist still examples where a
point of $M$ and one of its c-boundary $\partial M$ are not $T_2$-related (see Example \ref{ex1}). As in the case of the g- and b-
boundaries, one can get rid of this inconvenience by using the
minimal modification of the original topology in Theorems
\ref{prop2} and \ref{t1}. In the remainder of Section \ref{s4}, we
will develop in detail this modification, in order to check that
all the other desirable properties of the c-boundary are preserved
by this minimal modification.

First, we will introduce some basic notions. A non-empty
%\footnote{OJO con esto!! Errores en articulos previos.}
subset $P\subset M$ is called a {\em past set} if it coincides
with its past; i.e., $P=I^{-}(P):=\{p\in M: p\ll q\;\hbox{for
some}\; q\in P\}$. The {\em common past} of $S\subset M$ is
defined by $\downarrow S:=I^{-}(\{p\in M:\;\; p\ll q\;\;\forall
q\in S\})$. In particular, the past and common past sets are open.
A past set that cannot be written as the union of two proper past
sets
%\footnote{Y con esto!! Errores en articulos previos.},
%both of which are also past sets,
is called {\em indecomposable past} set, {\em IP}. An IP which
does coincide with the past of some point of the spacetime
$P=I^{-}(p)$, $p\in M$ is called {\em proper indecomposable past
set}, {\em PIP}. Otherwise, $P=I^{-}(\gamma)$ for some
inextendible future-directed timelike curve $\gamma$, and it is
called {\em terminal indecomposable past set}, {\em TIP}. The dual
notions of {\em future set}, {\em common future}, {\em IF}, {\em PIF} and {\em
TIF}, are defined just by interchanging the roles of
past and future in previous definitions.

To construct the {\em future} and {\em past c-completion}, first
we have to identify each {\em event} (point) $p\in M$ with its
PIP, $I^{-}(p)$, and PIF, $I^{+}(p)$. This is possible in any {\em
distinguishing} spacetime, that is, a spacetime which satisfies
that two distinct events $p, q$ have distinct chronological
futures and pasts ($p\neq q \Rightarrow I^\pm (p) \neq I^\pm
(q)$). In order to obtain consistent topologies in the
c-completions, we will focus on a somewhat more restrictive class
of spacetimes, the {\em strongly causal ones}. These are
characterized by the fact that the PIPs and PIFs constitute a
sub-basis for the topology of the manifold $M$.

%if any of its points is
%characterized by its past and future, {\em strongly causal} if it
%does not admit neither closed nor ``almost closed'' causal curves,
%

Now, every event $p\in M$ can be identified with its PIP,
$I^-(p)$. So, the {\em future c-boundary} $\hat{\partial}M$ of $M$
is defined as the set of all the TIPs in $M$, and  {\em the future
c-completion} $\hat{M}$ becomes the set of all the IPs:
\[
M\equiv \hbox{PIPs},\qquad \hat{\partial}M\equiv
\hbox{TIPs},\qquad\hat{M}\equiv \hbox{IPs}.
\]
Analogously, each $p\in M$ can be identified with its PIF,
$I^+(p)$. The {\em past c-boundary} $\check{\partial}M$ of $M$ is
defined as the set of all the TIFs in $M$, and  {\em the past
c-completion} $\check{M}$ is the set of all the IFs:
\[
M\equiv \hbox{PIFs},\qquad \check{\partial}M\equiv
\hbox{TIFs},\qquad\check{M}\equiv \hbox{IFs}.
\]

For the (total) c-boundary, the so-called S-relation comes into
play \cite{Sz}.
%First, we will identify $V$ with the subset of
%$\hat{V}\times \check{V}$ formed by all the pairs
%$(I^-(p),I^+(p))$.
Denote $\hat{M}_{\emptyset}=\hat{M}\cup \{\emptyset\}$ (resp.
$\check{M}_{\emptyset}=\check{M}\cup \{\emptyset\}$). The
S-relation $\sim_S$ is defined in $\hat{M}_{\emptyset}\times
\check{M}_{\emptyset}$ as follows. First, in the case $(P,F)\in
\hat{M}\times \check{M}$,
\begin{equation} \label{eSz}  P\sim_S F \Longleftrightarrow \left\{
\begin{array}{l}
P \quad \hbox{is included and is a maximal IP into} \quad
\downarrow F
 \\
F \quad \hbox{is included and is a maximal IF into} \quad \uparrow
P.
\end{array} \right.
\end{equation}
By {\em maximal} we mean that no other $P'\in\hat{M}$ (resp.
$F'\in \check{M}$) satisfies the stated property and includes
strictly $P$ (resp. $F$). Recall that, as proved by Szabados
\cite{Sz}, $I^-(p) \sim_S I^+(p)$ for all $p\in M$, and these are
the unique S-relations (according to our definition (\ref{eSz}))
involving proper indecomposable sets. Now, in the case $(P,F)\in
\hat{M}_{\emptyset}\times \check{M}_{\emptyset}\setminus
\{(\emptyset,\emptyset)\}$, we also put
\[ %\label{eSz2}
P\sim_S \emptyset, \quad \quad (\hbox{resp.} \; \emptyset \sim_S F
)\] if $P$ (resp. $F$) is a (non-empty, necessarily terminal)
indecomposable past (resp. future) set that  is not S-related by
(\ref{eSz}) to any other indecomposable set; notice that
$\emptyset$ is never S-related to itself.

Now, we can introduce the notion of c-completion, according to
\cite{FHS}:
\begin{definition}\label{def4}
Let $(M,g)$ be a strongly causal spacetime. The c-completion of
$(M,g)$ is defined as follows.
\begin{itemize}
\item As a point set: the {\em c-completion} $\overline{M}$ is
formed by the pairs of TIPs and TIFs which are S-related, that is,
\[
\overline{M}:=\{(P,F)\in \hat{M}_{\emptyset}\times
\check{M}_{\emptyset}: P\sim_S F\}.
\]
Every point $p\in M$ of the manifold will be identified with its
corresponding pair $(I^-(p),I^+(p))$, so $M$ will be considered a
subset of $\overline{M}$ and, thus, the {\em c-boundary} is
defined as $\partial M=\overline{M}\backslash M$. \item As a
chronological set: two pairs $(P,F),(P',F')\in \overline{M}$ are
chronologically related, denoted $(P,F)\overline{\ll}(P',F')$, if
$F\cap P'\neq\emptyset$. \item Topologically: $\overline{M}$ is
endowed with the {\em chronological topology} $\tcrono$, i.e., the
sequential topology associated to the following limit operator
(recall Definition \ref{def3}):
\[%\label{eq4}
\lcrono(\sigma):=\left\{(P,F)\in \overline{M}:\begin{array}{l}
P\in \hat{L}(\{P_n\}) \hbox{ if $P\neq \emptyset$}\\
F\in \check{L}(\{F_n\}) \hbox{ if $F\neq \emptyset$}
\end{array}
\right\}\qquad\hbox{for any $\sigma=\{(P_n,F_n)\}\subset
\overline{M}$,}
\]
where \[%\label{eq5}
\begin{array}{l}P\in \hat{L}(\{P_n\}) \iff P\subset {\rm LI}(\{P_n\}) \hbox{ and it is maximal in } {\rm LS}(\{P_n\})\\
F\in \check{L}(\{F_n\}) \iff F\subset {\rm LI}(\{F_n\}) \hbox{ and it is
maximal in } {\rm LS} (\{F_n\}).\end{array}
\]
\end{itemize}
\end{definition}

The main properties of this choice of c-completion
$(\overline{M},\overline{\ll},\tcrono)$ are summarized in the
following result  (see \cite[Theorem 3.27]{FHS}):
\begin{theorem}\label{theo1}
The c-completion $(\overline{M},\overline{\ll},\tcrono)$ in Defn.
\ref{def4} satisfies the following properties:
\begin{itemize}
\item[(i)] Any future-directed (resp. past-directed) inextendible
timelike curve in $M$ (namely, $\gamma: [a,b)\rightarrow M$) has
an endpoint in $\partial{M}$.
%{\color{red} In fact, if $\gamma$ is a future (resp. past) directed curve, for any sequence $t_n\nearrow \Omega$ we have that:
%\begin{equation}\label{eq2}
%\begin{array}{c}
%L(\{\gamma(t_n)\})=\{(P,F)\in \partial M:P=I^-(\gamma)\}\\
%\left(L(\{\gamma(t_n)\})=\{(P,F)\in \partial M:F=I^+(\gamma)\}
%  \right)
% \end{array}
%\end{equation}
%}
\item[(ii)] The inclusion $i:M\hookrightarrow \overline{M}$ is a
topological embedding and $i(M)$ is dense in $\overline{M}$.
\item[(iii)] The c-boundary $\partial M$ is closed in
$\overline{M}$. \item[(iv)] $I^\pm((P,F))$ is open for any
$(P,F)\in \overline{M}$. \item[(v)] The topology $\tau_{chr}$ is
$T_1$.
\end{itemize}
\end{theorem}

\subsection{The modified c-boundary}\label{s4.3}

%OLD VERSION: MY MAIN OBJECTION S THE FOOTNOTE WITH !!!

Given the limit operator $\lcrono$ and its associated chronological topology $\tcrono := \tau_{\lcrono}$ on $\overline{M}$, we put $D=M$ and consider the minimally $M$-separating topology $\tnew$, the minimally sequentially $M$-separating topology $\tnewL$, the operator $\lnew$ and its associated topology $\tau_{\lnew}$. In the beginning of Section \ref{s}, we imposed a
``minimality'' condition for our refinements in order to ensure that
both, the original topology and the refined ones shared as many
properties as possible (except the new properties). Therefore, it is expected that the
refinements defined along such a section (summarized in Remark
\ref{r} and Figure \ref{figresumen}) will also satisfy
the properties included in Theorem \ref{theo1}. In fact, we can
prove the following result:

%
%replace in the previous completion the chronological
%topology $\tau_{chr}$ by {\Cambios the three topologies provided in Section \ref{s} (see Remark \ref{remresumen} and Figure \ref{figresumen}).}
%%the unique \crust topology
%%$\tau_{chr}^{*}$ provided in Theorem \ref{prop2}.
% The next result
%shows  that all the  properties in Theorem \ref{theo1} are
%preserved by these new completions.
%\footnote{{\bf M.: OJO. Aqui'
%parece que no se ha tenido en cuenta nada de los u'ltimos avances.
%Os recuerdo que tenemos dos posibilidades: una es la topologia
%$\tau_{chr}^{*}$ y la otra es la topologi'a
%$\tau_{L_{\tau_{chr}^{*}}}$. Aqui se enuncia el teoremA para la
%primera, y estaria bien que se verificara asi. Sin embargo, lo
%necesitaremos realmente para la segunda, porque esa es la que
%aparecera' ma's abajo en el Corollary \ref{c} y Th. \ref{t}. POR
%TANTO ES IMPORTANTE REVISAR ESTO, COMPROBAR CON CUA'L DE LAS DOS
%TOPOLOGI'AS SE OBTIENE EL RESULTADO, Y TRATAR COMO SEA DE
%OBTENERLO PARA LA SEGUNDA}}

\begin{theorem}\label{theo2}
The topological spaces {\Cambios $(\overline{M},\tnew)$,
$(\overline{M},\tnewL)$ and
$(\overline{M},\tau_{\lnew})$}  endowed with the chronological relation $\overline{\ll}$ satisfy all the assertions (i)--(v) in Theorem \ref{theo1}.
%the following properties:
%\begin{itemize}
%\item[(i)] Any future-directed (resp. past-directed) inextendible
%timelike curve in $M$ (namely, $\gamma: [a,b)\rightarrow M$) has
%an endpoint in $\partial{M}$.
%%\item[(i)] Any {\Cambios inextendible} timelike curve in $M$ has
%%an endpoint in $\partial M$.
%\item[(ii)] The inclusion
%$i:M\hookrightarrow \overline{M}$ is a topological embedding and
%$i(M)$ is dense in $\overline{M}$. \item[(iii)] The c-boundary
%$\partial M$ is closed in $\overline{M}$. \item[(iv)]
%$I^\pm((P,F))$ is open for any $(P,F)\in\overline{M}$.
%\item[(v)]{\Cambios The topologies are $T_1$.}
%\end{itemize}
Moreover, the points on $M$ are $T_2$-separated from the
points on $\partial M$ with the topologies $\tnew$ and
$\tnewL$, and the latter is equal to $\tau_{\lnew}$ if $\lcrono$ is of first order.

\end{theorem}

\Demo Assertions (iii), (iv) and (v) are straightforward as all topologies are finer than the chronological one, and this one satisfies them (by Theorem \ref{theo1}). Next, recall that, for any future (resp. past) inextendible timelike curve $\gamma:[a,b)\rightarrow M$ and any sequence $t_{n}\nearrow b$, the following equation
\[
\begin{array}{c}
L(\{\gamma(t_n)\})=\{(P,F)\in \partial M:P=I^-(\gamma)\}\\
\left(\hbox{resp.}\;\, L(\{\gamma(t_n)\})=\{(P,F)\in \partial M:F=I^+(\gamma)\}
  \right)
 \end{array}
\]
is true for both, $\lcrono$ and $\lnew$. So, assertion (i) is implicitly proved in \cite[Theorem 3.27]{FHS}, while the second part of (ii) is a consequence of the fact that no sequence $\{\gamma(t_n)\}$ with $t_n\nearrow b$ converges to a point in $M$. On the other hand, the first assertion in (ii)
follows from the first assertion of Theorem \ref{theo1} (ii) and
the fact that $\lnew\mid_{M}=\lcrono\mid_{M}$ (recall \eqref{def2}, i.e., formula
(\ref{deflimchrmod}) below and Proposition \ref{lem2}). Finally, the last assertion is obtained from Theorem \ref{prop1}.\cvd

\begin{remark}{\em (1) The topology $\tnew$ may not be
Hausdorff, as two points at the boundary $\partial M$ may be
non-T$_2$-separated. However, such type of examples are natural in
different situations. Namely, this happens if one removes a half
time axis of Lorentz-Minkowski $\LL^2$ (as the removed origin
would yield naturally two boundary points), or in more refined
examples such as the ``grapefruit on a stick'' in \cite{FH} or the
two chimneys construction in \cite[Appendix]{FHSst}. In these
examples, the ideal points which are non-T$_2$-separated for
$\tcrono$ remain non-T$_2$-separated for $\tnew$
---but this can be regarded as harmless.

(2) Conditions (i)--(v) of both,  Theorem \ref{theo1} and
 Theorem \ref{theo2}, remain true with independence of the fact that
 $\lcrono$ or $\lnew$ may or not be of first
order. The first order condition is only necessary to ensure that $\tau_{\lnew}$ satisfies (\Asep).}
\end{remark}

\section{Admissibility criteria for c-boundaries} \label{s5}

A strong support for the choice of the explained definition of
{\em c-completion}  in \cite{FHS}, is that  such a choice follows
from a set of minimal hypotheses, which  catch the intuitive
requirements that a c-boundary must fulfill. But, as pointed out
in that reference, one could add more hypotheses if further
properties were required for that boundary. Let us revisit first
the original hypotheses, and then, let us add the minimal separability of the boundary as one of these
hypotheses.

\subsection{Original criteria}\label{s5.1}
The admissibility conditions for the c-boundary in \cite{FHS}
provide the point set and chronological structures previously
explained in subsection \ref{s4.2}. Let us remind those for
the topology.

\begin{definition}\label{def5} Consider a strongly causal spacetime $M$ and its c-completion
$\overline{M}$ regarded only as  a point set and a chronological
set (see  Defn. \ref{def4}). The following hypotheses are {\em
conditions of admissibility} for a  topology $\tau$ on
$\overline{M}$:
\begin{itemize}
\item[(A1)] For all $(P,F)\in \overline{M}$, the sets
$I^\pm((P,F))$ are open. \item[(A2)] The limits for $\tau$ are
compatible with the empty set, i.e., if
$\{(P_n,F_n)\}_n\rightarrow (P,\emptyset)$ (resp. $(\emptyset,F)$)
and there exists $(P',F')\in \overline{M}$ such that $P\subset
P'\subset {\rm LI}(\{P_n\})$ (resp. $F\subset F'\subset{\rm
LI}(\{F_n\})$), then $(P',F')=(P,\emptyset)$ (resp.
$(P',F')=(\emptyset,F)$).

\item[(\AminM)] $\tau$ is minimally fine
among the topologies satisfying previous conditions, i.e., no
other topology satisfying the conditions (A1) and (A2) is strictly
coarser than $\tau$.

\item[(\AminML)] $\tau$ is
minimally fine among the sequential topologies satisfying previous
conditions (A1) and (A2), i.e., $\tau$ is sequential and no other
sequential topology satisfying conditions (A1) and (A2) is
strictly coarser than $\tau$.
\end{itemize}
A topology $\tau$ satisfying the conditions (A1), (A2) and
(\AminM) will be called {\em generally
admissible} and one satisfying (A1), (A2) and
 (\AminML) will be {\em
 sequentially admissible} or simply {\em admissible}. Accordingly,
 the c-completion $\overline{M}$ endowed with such a topology will
 be called {\em generally admissible} or {\em admissible}.
\end{definition}

%END END END
\begin{remark}{\rm
 Here we have reserved the term ``admissible topology''
for the plain notion defined above, which agrees with \cite{FHS} and subsequent references.
%In the original definition (see \cite[Sect. 3.1.3]{FHS}) the
%authors considered the possibility to include some additional
When, in the spirit of \cite[Sect. 3.1.3]{FHS}, further hypotheses
are imposed, we will use a different name  (as in Definition
\ref{def1} below), in order to avoid confusions.}
\end{remark}

\noindent Let us discuss briefly the conditions in Definition
\ref{def5} (see \cite{FHS} for further details). Condition (A1)
determines partially the convergence of sequences. In fact, for
any topology $\tau$ satisfying (A1):
\[
(P_n,F_n)\rightarrow (P,F) \Rightarrow P\subset {\rm LI}(\{P_n\}),\;
F\subset {\rm LI}(\{F_n\}).
\]
Moreover, if, in addition, $P\neq\emptyset\neq F$, we have the following implication in terms of the operator $\lcrono$ of the chr-chronology:
\begin{equation}\label{eq6}
(P_n,F_n)\rightarrow (P,F) \Rightarrow (P,F)\in
\lcrono(\{(P_n,F_n)\}).
\end{equation}
(see \cite[Lemma 3.15]{FHS} for details). Condition (A2) is a compatibility requirement for the convergence
of sequences when a component of the limit pair is empty.
{\Cambios In particular,
one can extend formula \eqref{eq6} to any $(P,F)\in \overline{M}$,
that is, } if a topology $\tau$ satisfies (A1) and (A2) and
$(P_n,F_n)\rightarrow (P,F)$ one has:
\[%begin{equation}\label{eq6'}
\begin{array}{c}
(P_n,F_n)\rightarrow (P,\emptyset) \Rightarrow (P,\emptyset)\in
\lcrono(\{(P_n,F_n)\}) \\ (\hbox{resp.}\;\;(P_n,F_n)\rightarrow
(\emptyset,F) \Rightarrow (\emptyset,F)\in \lcrono(\{(P_n,F_n)\})).
\end{array}
\]%end{equation}
(see \cite[Prop. 3.20]{FHS} for details). Summing up, the following property is obtained for  referencing:
\begin{lemma}\label{ll'} If $\sigma\rightarrow (P,F)$ with any topology $\tau$ satisfying (A1), (A2) then $(P,F)\in L_{chr}(\sigma)$.
\end{lemma}
Finally, conditions ($\overline{\rm A}_{\scriptsize Min}$) or ($\overline{\rm A}^{\scriptsize Seq}_{\scriptsize Min}$) are
minimality conditions which will allow to speak on uniqueness and guarantee that no spurious sets are
included in the topology.

From the viewpoint of ``first principles'', the following result,
which clarifies \cite[Theorem 3.22]{FHS}, justifies the
choice of the chronological topology for the c-completion (at
least when $\lcrono$ is of first order).
\begin{theorem}\label{theo5}
Let $\overline{M}$ be the c-completion  of a strongly causal
spacetime $M$ regarded as a point set and a chronological set. If the limit operator $\lcrono$ is of first order,
then the chr-topology is the unique (sequentially) admissible
topology on $\overline{M}$. Moreover, in this case,  $\lcrono$ is also the associated limit
operator of any generally admissible topology $\tau\subset \tcrono$.
\end{theorem}

\Demo The first assertion is obtained in \cite[Theorem
3.22]{FHS}, but we can sketch now a proof by using our previous
results as follows. First, a straightforward  computation shows
that $\lcrono$ satisfies (A1) and (A2). By Lemma \ref{ll'},
the limit operator $L$ associated to any sequential topology
obeying (A1) and (A2) must satisfy $L\subset \lcrono$, and
the minimality assumption ($\overline{\rm A}_{\scriptsize Min}^{\scriptsize Seq}$) implies $L=\lcrono$.

For the last assertion, assume that there exists a generally admissible topology $\tau\subset\tcrono$ and denote by $L_{\tau}$ its associated limit operator. On the one hand, from Proposition \ref{p} (b), we deduce that $\tau_{Seq}\subset \tau_{\lcrono}$ and, as $L_{\tau}$ is of first order, Lemma \ref{lem3.3a} ensures that $\lcrono(\sigma)\subset L_{\tau}(\sigma)$ for all sequence $\sigma\subset \overline{M}$. On the other hand, recall that the limit operator $L_{\tau}$ is associated to a topology $\tau$ fulfilling the admissibility conditions, and so, it must also obey $L_\tau(\sigma)\subset \lcrono(\sigma)$ for all sequence
by Lemma \ref{ll'}. In conclusion, $L_{\tau}=\lcrono$. \cvd
%
%recall first that the limit operator
%$L_\tau$ associated to any topology $\tau$ fulfilling the three
%admissibility conditions, must also obey $L_\tau\subset \lcrono$
%by Lemma \ref{ll'}. So, if $L_\tau\neq \lcrono$ then $\tau$ is not
%coarser than $\lcrono$. Now, consider the set $\mathcal{S}$ of all
%the topologies with associated limit operator equal to $\lcrono$
%(recall Remark \ref{r2.5}(1)). This set is partially ordered by
%the binary relation $<:=$ ``is finer than''. Moreover,
%$(\mathcal{S},<)$ is an inductive set\footnote{!!! CHECK! If this
%were not true, we could not ensure existence (but if it exists,
%the remainder holds) } and, so, Zermelo's lemma ensures the
%existence of the required maximal elements. \cvd
%(ii) As $\lcrono$ is not of first order, the  chr-topology will
%satisfy $\sigma\rightarrow (P,F)$ but  $(P,F)\not\in
%L_{chr}(\sigma)$ for some sequence and, so, by Lemma \ref{ll'} it
%will not satisfy the conditions of admissibility. Then, we focus
%in the set $\mathcal{S}'$ of all the sequential topologies (resp.
%topologies) $\tau$ on $\overline{M}$ whose associated limit
%operator $L_\tau$ satisfies $L_\tau\subset \lcrono$. Recall that
%$\mathcal{S}'$ is not empty, as it contains the topology $\tau_0$
%which admits as topological basis the natural topology of $M$
%union the discrete topology on $\partial M$. Considering  the
%binary relation $<$ and reasoning as above,\footnote{!!! I have
%even more doubts than before on the applicability of Zermelo's
%lemma here.} any maximal element of $(\mathcal{S}',<)$ is the
%required topology. \cvd

\smallskip

%(END OF NEW REDACTION)

\begin{remark} \label{r0} {\em
As  the operator $\lcrono$ is not of first  order only in
very artificial cases (see Example \ref{eluis}), this theorem
(plus the properties of Theorem \ref{theo1}) gives a strong
support for the usage of the chr-topology, at least if one does not care on the $T_2$-separability of the boundary. In fact, if a redefinition of
the c-boundary topology preserved the expected good properties,
then it would agree with the chr-topology whenever $\lcrono$ is of
first order.
% We cannot ensure in general the existence of generally admissible topologies. However, from
Moreover, in this case the last assertion of previous theorem ensures that, whenever generally admissible topologies exist, they are minimally fine among the topologies with $\lcrono$ as an associated limit. When $\lcrono$ is not of first order then the chronological topology is not admissible (this can be checked directly from Lemma \ref{ll'}). However, the chronological topology would still make sense and 
%so, this property will not be imposed a priori.}
we will explore even this case.} \end{remark}

\subsection{Including $D$-separability as an admissibility condition}\label{s5.2}
As shown in Section \ref{s}, the $T_2$-separation of points of $M$
and $\partial M$ can be always obtained as a requirement {\em a
posteriori}. However, one can think that such a  property
 should be included as one of the  {\em a priori} properties of the c-boundary,
 at the same level of the other admissibility conditions.
Next, this approach is developed.

Recall first that the limit operator of the searched
($T_2$-separating) topology $\tau$ must satisfy also the
admissibility conditions (A1) and (A2) and, so, Lemma \ref{ll'}
implies $L_\tau \subset \lcrono$. We focus on the case of
sequential topologies because, on the one hand, the topology of the c-boundary is studied by considering the convergence of some sequences and, on the other, results such as Theorem 5.4 show that they are specially interesting (of course, this can be extended to non-sequential ones, as we have done till now).

\begin{definition}\label{def1}
A sequential topology $\tau$ is {\em \tadmisible} for the
c-completion $\overline{M}$ if it satisfies properties (A1) and
(A2) of Definition \ref{def5}, together with (\Asep) in Definition \ref{dcrust} 
%which reads here:
%\begin{itemize}
%
%\item[(\Asep)] T$_2$-separability of points
%of $M$ and $\partial M$; i.e., for every $p\in M$ and every
%$(P,F)\in
%\partial M$, there exists an open set $U$ of $M$ and some
%$V\in\tau$ such that $p\in U$, $(P,F)\in V$ and $U\cap
%V=\emptyset$.
%\end{itemize}
and the following one:

\begin{itemize}
	
\item[($\tilde{A}_{\scriptsize Min}^{\scriptsize Seq}$)] $\tau$ is minimally fine among the sequential topologies
satisfying previous conditions, i.e., no other sequential topology
satisfying (A1), (A2) and (\Asep) is
strictly coarser than $\tau$.
\end{itemize}
\end{definition}
Taking into account Proposition \ref{prop1'}, we consider the
operator
\begin{equation}\label{deflimchrmod}
\lnew(\sigma):=\left\{\begin{array}{lc}\lcrono(\sigma)\cap M &
\hbox{if $\exists\, \kappa\subset \sigma$ and $p\in M$ such that }
p\in \lcrono(\kappa)  \\ \lcrono(\sigma) & \hbox{otherwise}
\end{array}\right.
\end{equation}
consistently with \eqref{def2}. If we apply Theorem \ref{prop1} (ii) to
$(\overline{M},\tcrono)$ and $D=M$, we deduce directly:
\begin{corolario}\label{c} Let $\overline{M}$ be the c-completion of a strongly causal spacetime
$M$ regarded as a point set and a chronological set. If  the limit operator $\lcrono$ is of first order, 
%on
%$\overline{M}$, 
then the operator $\lnew$ is also of first
order, and its associated  topology $\tau_{\lnew}$ is equal to
$\tnewL$, i.e., $\tau_{\lnew}$ is the
unique sequentially minimally $D$-separating topology for $D=M$.
\end{corolario}
However, one can go a step further in order to check that
$\tau_{\lnew}$ is the suitable topology  for the requirements
in Definition \ref{def1} (in the same sense that Theorem
\ref{theo5} proved that $\tau_{\lcrono}$ was the suitable topology
for Definition \ref{def5}) by imposing directly the first order character to $\lnew$.

\begin{theorem}\label{t} Let $\overline{M}$ be the c-completion of a strongly causal spacetime
$M$ regarded as a point set and a chronological set. If $\lnew$ is of first order on $\overline{M}$, then the topology $\tau_{\lnew}$ satisfies (A1), (A2) and (\Asep). If, in addition,
$\lcrono$ is of $k$-th order for some ordinal $k$ (in particular, both hypotheses hold when $\lcrono$ is of first order), then
$\tau_{\lnew}$ also satisfies ($\tilde{A}^{\scriptsize Seq}_{\scriptsize Min}$), i.e., $\tau_{\lnew}$ is the unique $T_2$-admissible sequential
topology of $\overline{M}$.
%satisfying (A1), (A2),
%(A$_{\hbox{\scriptsize Sep}}$) and (A$_{\hbox{\scriptsize
%Min}}^{\hbox{L}}$).
\end{theorem}

\Demo Let us begin with the first assertion, that is, let us show that $\tau_{\lnew}$ satisfies (A1), (A2) and (\Asep) if $\lnew$ is of first order. Since $\lnew\subset L_{chr}$, thus $\tcrono\subset
\tau_{\lnew}$ (see Lemma \ref{lem3.3a}), and $\tcrono$
satisfies (A1), we deduce that $\tnew$ also satisfies (A1).
Moreover, from Proposition \ref{prop1'} and Lemma \ref{lem3.3b},
we know that $\tau_{\lnew}$ satisfies (\Asep). In order to prove (A2), assume that
$\sigma=\{(P_n,F_n)\}\rightarrow (P,\emptyset)$, $P\neq\emptyset$, with
$\tnew$ (the case $P=\emptyset\neq F$ is analogous)
%\end{document}
and assume the existence of $(P',F')\in\overline{M}$ such that
$P\subset P'\subset {\rm LI}(\{P_n\})$. As $\lnew$ is of first
order on $\overline{M}$, $(P,\emptyset)\in \lnew(\sigma)$.
Taking into account that $(P,\emptyset)\in\partial M$, necessarily
$\lnew(\sigma)=\lcrono(\sigma)$, and thus, $(P,\emptyset)\in
\lcrono(\sigma)$. In particular, $P$ is a maximal IP in ${\rm
LS}(\{P_n\})$. Therefore, as $(P',F')\in\overline{M}$ satisfies
$P\subset P'\subset {\rm LI}(\{P_n\})$, necessarily $P=P'$ and
$F'=\emptyset$, as required.

{\Cambios For the second assertion, assume additionally that $\lcrono$ is of $k$-th order and let us show that $\tau_{\lnew}$ also satisfies (\ALMin). In fact, we are going to prove something even stronger, concretely, that any sequential topology
$\tau_{L}$ satisfying (A1), (A2) and (\Asep), must obey  $\tau_{\lnew}\subset \tau_{L}$. On the one hand, as $\lcrono$ is of $k$-th order, Proposition \ref{prop4.2} ensures that $\lnew=L^*_{\tcrono}$, and so, from Theorem \ref{prop1} (ii), we deduce that $\tau_{\lnew}$ is the unique sequentially minimally $D$-separating topology. On the other hand, as $\tau_{L}$ satisfies (A1) and (A2), Lemma \ref{ll'} ensures that  $L(\sigma)\subset \lcrono(\sigma)$ for all sequence $\sigma$ and so, $\tcrono\subset \tau_{L}$, i.e., $\tau_{L}$ satisfies (\AFin). Taking into account that $\tau_{L}$ also satisfies (\Asep) and the definition of sequentially minimally $D$-separating topology, we deduce that $\tau_{\lnew}\subset \tau_{L}$, as desired.  \cvd

\smallskip

Figure \ref{figresumen2} summarizes the results of this section, for the
convenience of the reader.

\begin{remark}\label{g} {\em
%As the hypothesis of being of first order is less restrictive when
%it is imposed on $\lcrono^*$ than on $\lcrono$, Theorem \ref{t}
%becomes more refined than Corollary \ref{c}.
As the hypotheses of Theorem \ref{t} are less restrictive than the first order property for $\lcrono$, its conclusion
is sharper than Corollary \ref{c} (see Example \ref{eluis}). But, beyond this
subtlety, either Theorem \ref{t} or Corollary  \ref{c} show that no
matter if the $T_2$-separability is imposed a priori ($T_2$-admissible)
or a posteriori ($D$-separating).

Summing up, the question of $T_2$-separability can be circumvented for the c-boundary. From a fundamental viewpoint, one
can work with the $T_2$-separating topology $\tau_{\lnew}$, as
the admissibility properties of this topology yield more accurate consequences than those for
 $\tau_{\lcrono}$ (Corollary \ref{c}, Theorem \ref{t}). In any case, as pointed out in Remark \ref{r0}, the cases
where the topologies $\tcrono$ or $\tau_{\lnew}$
satisfy the conditions of admissibility are so general that one
would not be especially worried with the cases where they differ. And, from
the practical viewpoint, this way of proceeding is equivalent to
consider the chr-completion (as defined in \cite{FHS}) and impose
the $T_2$-separation of $M$ and $\partial M$, as one can do in the
general framework of Section \ref{s}. Nevertheless, if further
studies suggested that more separability properties must be
required for the c-boundary, one could remake our previous process
by including such properties, in the spirit of \cite{FHS}.}
\end{remark}

\begin{figure}
\centering \ifpdf
  \setlength{\unitlength}{1bp}%
  \begin{picture}(438.66, 446.50)(0,0)
  \put(0,0){\includegraphics{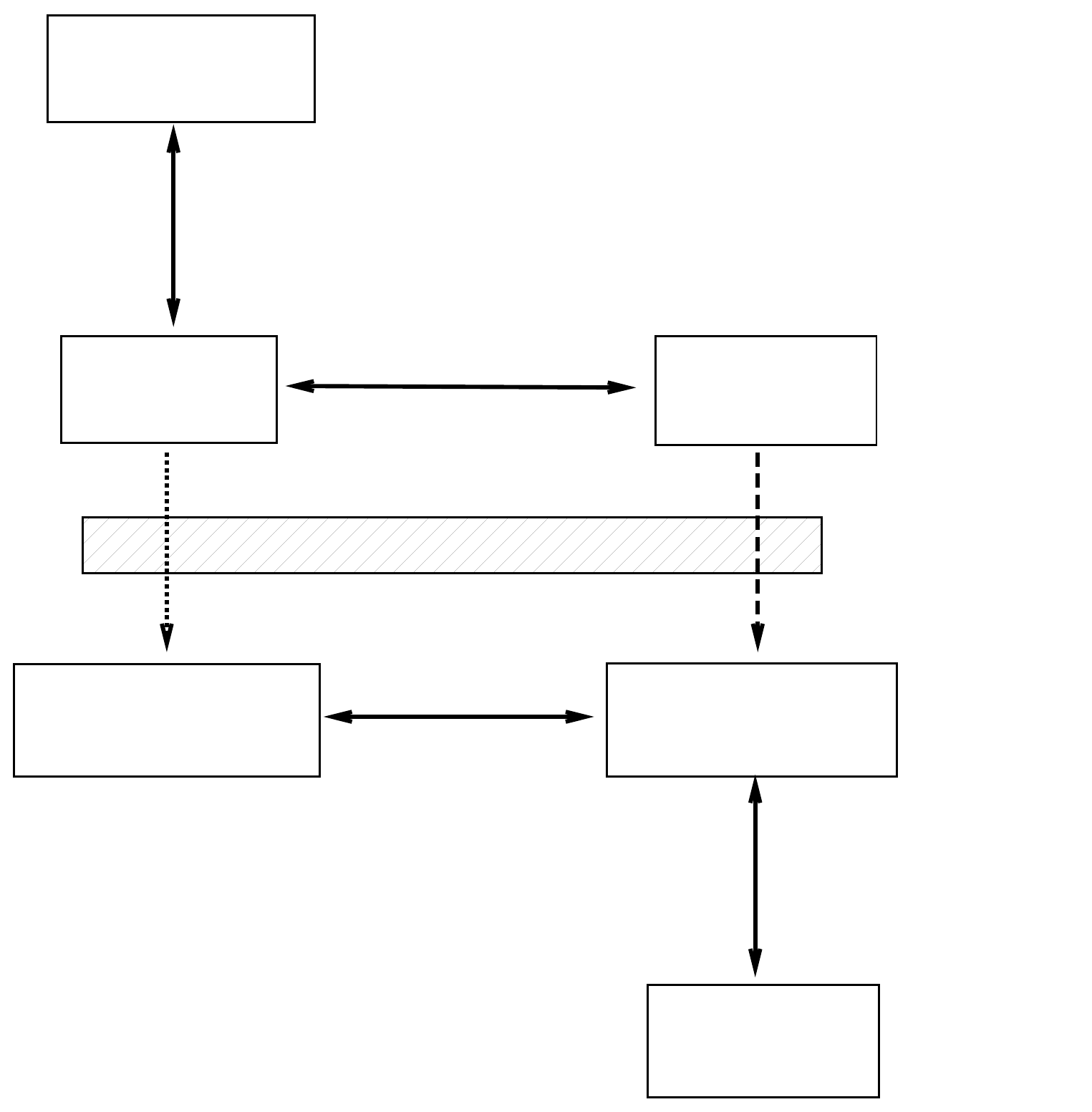}}
  \put(48.63,425.88){\fontsize{8.83}{10.60}\selectfont Admisibility }
  \put(36.62,294.92){\fontsize{8.83}{10.60}\selectfont  Chronological}
  \put(39.62,280.92){\fontsize{8.83}{10.60}\selectfont topology $\tcrono$}
  \put(268.34,295.04){\fontsize{8.83}{10.60}\selectfont  Chronological limit}
  \put(282.34,280.04){\fontsize{8.83}{10.60}\selectfont operator $\lcrono$}
  \put(26.36,409.13){\fontsize{8.83}{10.60}\selectfont (A1), (A2) and (\ALMin)}
  \put(112.21,223.82){\fontsize{8.83}{10.60}\selectfont Refining the topology with $D=M$}
  \put(75.33,362.94){\fontsize{8.83}{10.60}\selectfont When $\lcrono$ is of first order}
  \put(312.39,198.28){\fontsize{8.83}{10.60}\selectfont Formula (\ref{deflimchrmod})}
  \put(153.58,297.03){\fontsize{8.83}{10.60}\selectfont Definition \ref{def4}}
  \put(140.11,146.46){\fontsize{8.83}{10.60}\selectfont When $\lcrono$ of first order}
  \put(250.79,157.15){\fontsize{8.83}{10.60}\selectfont $\lnew\rightarrow$ derived top. $\tau_{\lnew}$}
  \put(96.33,350.14){\fontsize{8.83}{10.60}\selectfont (Theorem \ref{theo5})}
  \put(273.451,35.73){\fontsize{8.83}{10.60}\selectfont $T_2$-admissibility:}
  \put(269.48,23.48){\fontsize{8.83}{10.60}\selectfont (A1), (A2), (\Asep) }
  \put(282.71,11.23){\fontsize{8.83}{10.60}\selectfont and (\ALMin)}
  \put(156.84,134.22){\fontsize{8.83}{10.60}\selectfont (Theorem \ref{prop1})}
  \put(308.77,106.16){\fontsize{8.83}{10.60}\selectfont When $\lcrono$ is of first order (Corollary \ref{c})}
  \put(308.77,92.16){\fontsize{8.83}{10.60}\selectfont or more generally $\lcrono^*$ of first order and}
  \put(308.77,77.16){\fontsize{8.83}{10.60}\selectfont  $\lcrono$ of $k$-th order (Theorem \ref{t})}
  \put(18.41,162.94){\fontsize{8.83}{10.60}\selectfont $\tnew \rightarrow$ limit op. $L_{\tnew}\rightarrow$}
  \put(19.14,147.79){\fontsize{8.83}{10.60}\selectfont derived topology $\tnewL$}
%  \put(243.77,144.11){\fontsize{8.83}{10.60}\selectfont derived topology $\tau_{\lnewd_{chr}}$}
  \put(75.57,198.17){\fontsize{8.83}{10.60}\selectfont Theorem \ref{prop2} (recall (\ref{eq11}))}
  \end{picture}%
\else
  \setlength{\unitlength}{1bp}%
  \begin{picture}(438.66, 446.50)(0,0)
    \put(0,0){\includegraphics{esquema2.pdf}}
    \put(48.63,425.88){\fontsize{8.83}{10.60}\selectfont Admisibility }
    \put(36.62,294.92){\fontsize{8.83}{10.60}\selectfont  Chronological}
    \put(39.62,280.92){\fontsize{8.83}{10.60}\selectfont topology $\tcrono$}
    \put(268.34,295.04){\fontsize{8.83}{10.60}\selectfont  Chronological limit}
    \put(282.34,280.04){\fontsize{8.83}{10.60}\selectfont operator $\lcrono$}
    \put(26.36,409.13){\fontsize{8.83}{10.60}\selectfont (A1), (A2) and (\ALMin)}
    \put(112.21,223.82){\fontsize{8.83}{10.60}\selectfont Refining the topology with $D=M$}
    \put(75.33,362.94){\fontsize{8.83}{10.60}\selectfont When $\lcrono$ is of first order}
    \put(312.39,198.28){\fontsize{8.83}{10.60}\selectfont Formula (\ref{deflimchrmod})}
    \put(153.58,297.03){\fontsize{8.83}{10.60}\selectfont Definition \ref{def4}}
    \put(140.11,146.46){\fontsize{8.83}{10.60}\selectfont When $\lcrono$ of first order}
    \put(250.79,157.15){\fontsize{8.83}{10.60}\selectfont $\lnew\rightarrow$ derived top. $\tau_{\lnew}$}
    \put(96.33,350.14){\fontsize{8.83}{10.60}\selectfont (Theorem \ref{theo5})}
    \put(273.451,35.73){\fontsize{8.83}{10.60}\selectfont $T_2$-admissibility:}
    \put(269.48,23.48){\fontsize{8.83}{10.60}\selectfont (A1), (A2), (\Asep) }
    \put(282.71,11.23){\fontsize{8.83}{10.60}\selectfont and (\ALMin)}
    \put(156.84,134.22){\fontsize{8.83}{10.60}\selectfont (Theorem \ref{prop1})}
    \put(308.77,106.16){\fontsize{8.83}{10.60}\selectfont Theorem \ref{t} when $\lnew$  of first order}
    \put(308.77,93.16){\fontsize{8.83}{10.60}\selectfont and $\lcrono$ of $k$-th order for some ordinal $k$}
    \put(308.77,77.16){\fontsize{8.83}{10.60}\selectfont (in particular if $\lcrono$ is of first order)}

    \put(18.41,162.94){\fontsize{8.83}{10.60}\selectfont $\tnew \rightarrow$ limit op. $L_{\tnew}\rightarrow$}
    \put(19.14,147.79){\fontsize{8.83}{10.60}\selectfont derived topology $\tnewL$}
  %  \put(243.77,144.11){\fontsize{8.83}{10.60}\selectfont derived topology $\tau_{\lnewd_{chr}}$}
    \put(75.57,198.17){\fontsize{8.83}{10.60}\selectfont Theorem \ref{prop2} (recall (\ref{eq11}))}
  \end{picture}%
\fi \caption{\label{figresumen2} This figure summarizes the main
results in Section \ref{s5}. \newline Recall that Section \ref{s}
provides two different ways to refine the topology $\tcrono$ in
order to obtain the separability condition on $D=M$: (a) focusing
on the chronological topology itself (following the dotted arrow)
or (b) considering the limit operator $\lcrono$ (following the
dashed arrow). When $\lnew$ is of first order the procedure
(b) is equivalent to include $T_2$-separability as an
admissibility condition for the c-boundary (Theorem \ref{t}). In
any case, when $\lcrono$ is of first order (which is slightly more restrictive than
$\lnew$ of first order), the procedures (a) and (b) are
equivalent (Theorem \ref{prop1}).}
\end{figure}

\section{Appendix: some examples}

This section is devoted to present some examples in order to illustrate some of the assertions appeared in previous sections. The first two examples were already studied in \cite{FHe} and \cite{A}, resp., so here we will just summarize the key points.

\begin{example}\label{ex1}{\em
Let us begin by showing the possibility for the c-boundary of the setting that motivates this paper, that is, the existence of a c-completion $\overline{M}$ where two points, one on the manifold and other on the boundary, are not T$_2$-related. For this, we will just recall the two examples introduced in \cite[Section 2.3]{FHe}. Such examples show  not only that the c-completion may present separability problems, but also that the involved boundary points can be represented by any type of pairs $(P,F)$  (i.e., with both or just one non-empty component).

On the first example (represented in Figure \ref{fig4}),
the non-empty sets $P$ and $F$ are S-related, and so, they determine a point at the boundary. Moreover, from the definition of the chronological limit (recall Def. \ref{def4}), we have that the sequence $\{p_n\}$ converges to both, the manifold point $p\in M$ and the boundary point $(P,F)\in\partial M$.

On the second example (Figures \ref{fig5} and \ref{fig6}), there are two boundary points attached to $q$. In fact, the set $P$ (which includes $P'$ but it is indecomposable) is S-related with $F$ while the set $P'$ is S-related with the empty set, obtaining the boundary points $(P,F),(P',\emptyset)\in\partial M$. Here, the sequence $\{p_n\}$ converges again to both, a point in the manifold $p\in M$, and a boundary point {\em having one empty component} $(P',\emptyset)\in \partial M$.
}
\begin{figure}
\centering
\ifpdf
  \setlength{\unitlength}{1bp}%
  \begin{picture}(266.81, 204.71)(0,0)
  \put(0,0){\includegraphics{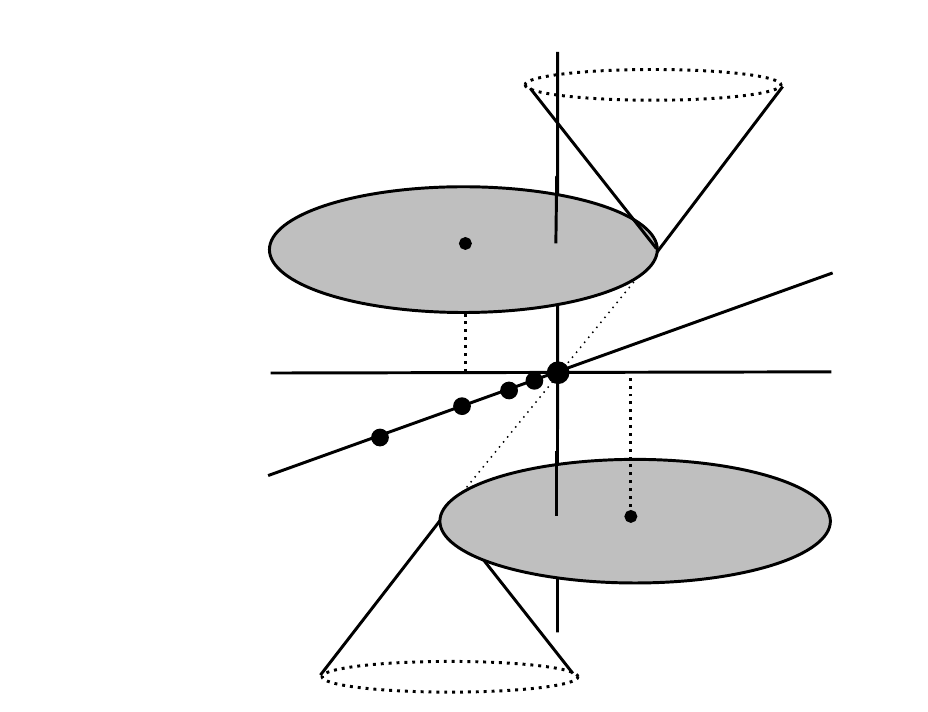}}
  \put(162.04,187.93){\fontsize{14.23}{17.07}\selectfont $t$}
  \put(226.08,87.57){\fontsize{14.23}{17.07}\selectfont $x$}
  \put(180.05,159.75){\fontsize{14.23}{17.07}\selectfont $F$}
  \put(119.60,24.27){\fontsize{14.23}{17.07}\selectfont $P$}
  \put(76.35,58.66){\fontsize{14.23}{17.07}\selectfont $y$}
  \put(108.35,67.66){\fontsize{11.38}{13.66}\selectfont $p_n$}
  \put(163.12,87.84){\fontsize{11.38}{13.66}\selectfont $p$}
  \end{picture}%
\else
  \setlength{\unitlength}{1bp}%
  \begin{picture}(266.81, 204.71)(0,0)
  \put(0,0){\includegraphics{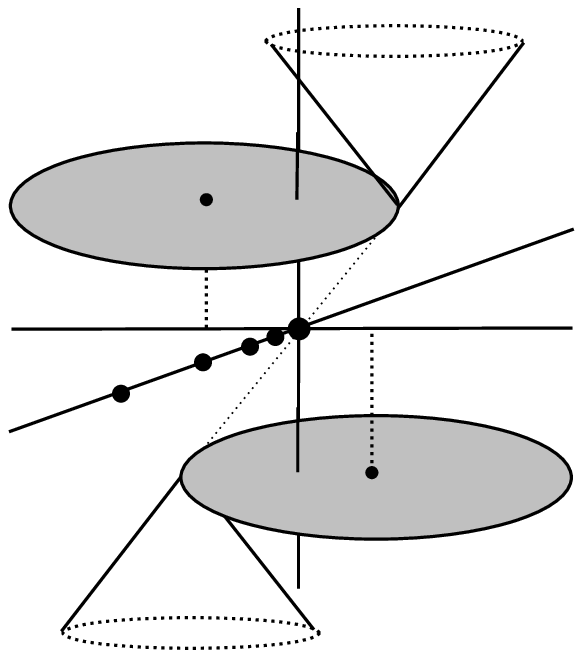}}
  \put(162.04,187.93){\fontsize{14.23}{17.07}\selectfont $t$}
  \put(226.08,87.57){\fontsize{14.23}{17.07}\selectfont $x$}
  \put(180.05,159.75){\fontsize{14.23}{17.07}\selectfont $F$}
  \put(119.60,24.27){\fontsize{14.23}{17.07}\selectfont $P$}
  \put(76.35,58.66){\fontsize{14.23}{17.07}\selectfont $y$}
  \put(161.12,87.84){\fontsize{11.38}{13.66}\selectfont $p$}
  \end{picture}%
\fi
\caption{\label{fig4}Minkowski spacetime $\LL^3$ with the dashed regions (two discs of radius $2$ centered at the points $(-1,0,1)$ and $(1,0,-1)$) removed.
%Here, the sets $P$ and $F$ are S-related and the sequence $\{p_n\}$ converges to both, $p\in M$ and $(P,F)\in \partial M$ with the chr. topology.}
}
\end{figure}

\begin{figure}
\centering \ifpdf
  \setlength{\unitlength}{1bp}%
  \begin{picture}(337.44, 175.37)(0,0)
  \put(0,0){\includegraphics{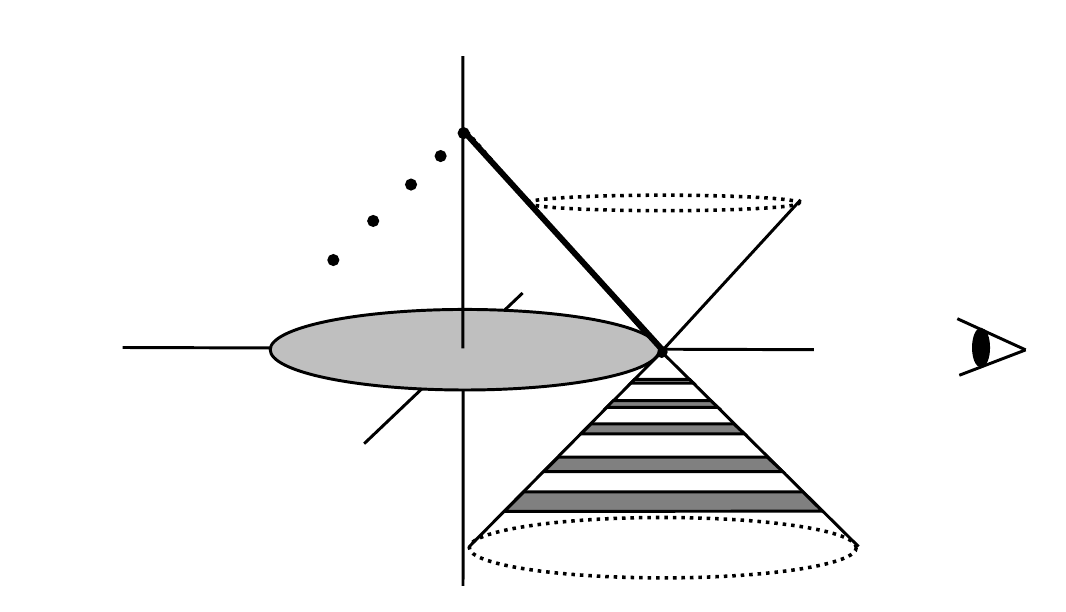}}
  \put(136.55,158.59){\fontsize{14.23}{17.07}\selectfont t}
  \put(236.62,69.93){\fontsize{14.23}{17.07}\selectfont x}
  \put(146.86,98.74){\fontsize{14.23}{17.07}\selectfont y}
  \put(139.03,137.69){\fontsize{14.23}{17.07}\selectfont p}
  \put(199.01,68.08){\fontsize{14.23}{17.07}\selectfont q}
  \put(95.25,119.17){\fontsize{14.23}{17.07}\selectfont $p_{n}$}
  \put(189.05,126.93){\fontsize{14.23}{17.07}\selectfont F}
  \put(156.26,8.73){\fontsize{14.23}{17.07}\selectfont P'}
  \put(201.35,20.40){\fontsize{14.23}{17.07}\selectfont P}
  \put(225.44,43.94){\fontsize{14.23}{17.07}\selectfont $\Pi_{n}$}
  \put(101.26,79.27){\fontsize{14.23}{17.07}\selectfont D}
  \end{picture}%
\else
  \setlength{\unitlength}{1bp}%
  \begin{picture}(337.44, 175.37)(0,0)
  \put(0,0){\includegraphics{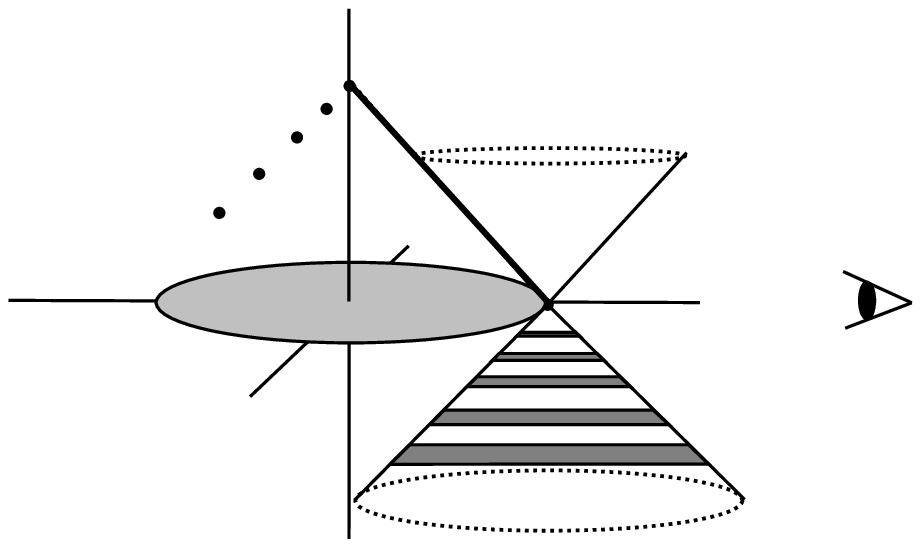}}
  \put(136.55,158.59){\fontsize{14.23}{17.07}\selectfont t}
  \put(236.62,69.93){\fontsize{14.23}{17.07}\selectfont x}
  \put(146.86,98.74){\fontsize{14.23}{17.07}\selectfont y}
  \put(139.03,137.69){\fontsize{14.23}{17.07}\selectfont p}
  \put(199.01,68.08){\fontsize{14.23}{17.07}\selectfont q}
  \put(95.25,116.17){\fontsize{8.54}{10.24}\selectfont $p_{n}$}
  \put(189.05,126.93){\fontsize{14.23}{17.07}\selectfont F}
  \put(156.26,8.73){\fontsize{14.23}{17.07}\selectfont P'}
  \put(201.35,20.40){\fontsize{14.23}{17.07}\selectfont P}
  \put(225.44,43.94){\fontsize{14.23}{17.07}\selectfont $\Pi_{n}$}
  \put(101.26,79.27){\fontsize{14.23}{17.07}\selectfont D}
  \end{picture}%
\fi\caption{\label{fig5}Representation of the spacetime $M$, obtained by
removing from $\LL^{3}$ the dashed regions, which consist of the
unitary disc $D$ and the sheets $\Pi_{n}$ (see Figure \ref{fig6}  for a lateral view of the sheets, which allows to distinguish between $P$ and $P'$).
%The figure also
%illustrates the sequence $\sigma=\{p_{n}\}_{n}$, which chr.
%converges to both, $p\in M$ and $(P',\emptyset)\in \partial M$.}
}
\end{figure}

\begin{figure}
\centering \ifpdf
  \setlength{\unitlength}{1bp}%
  \begin{picture}(336.87, 193.36)(0,0)
  \put(0,0){\includegraphics{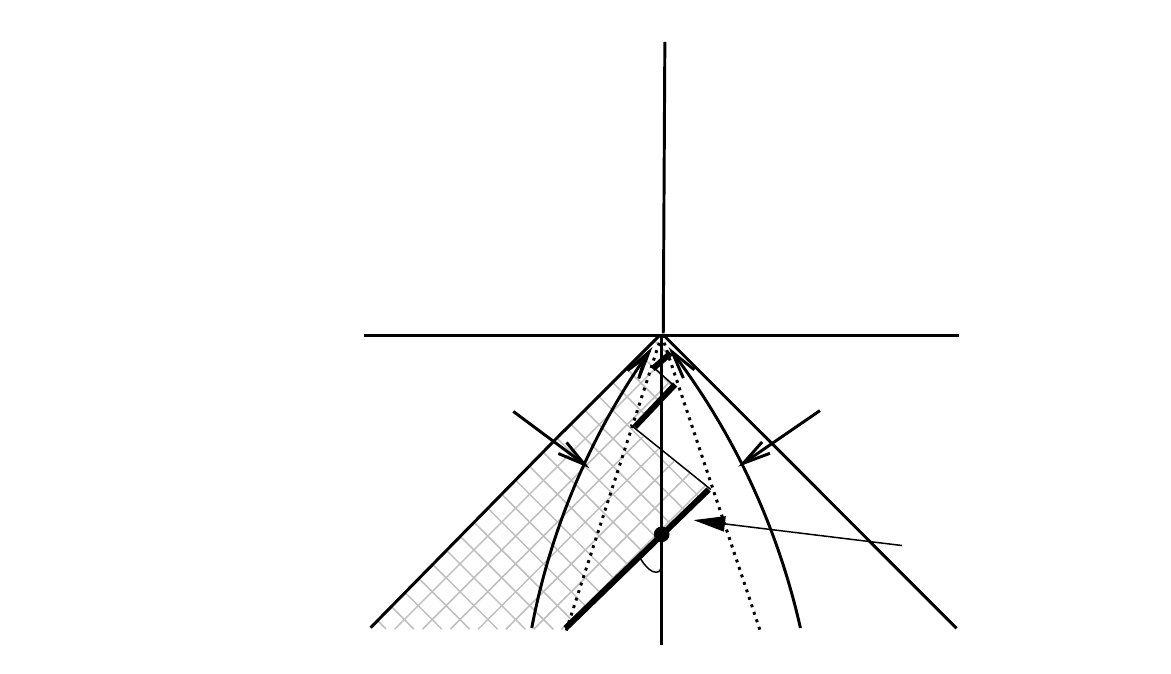}}
  \put(279.11,88.97){\fontsize{14.23}{17.07}\selectfont y}
  \put(196.23,176.58){\fontsize{14.23}{17.07}\selectfont t}
  \put(261.69,32.08){\fontsize{14.23}{17.07}\selectfont $L_{n}$}
  \put(170.56,8.73){\fontsize{14.23}{17.07}\selectfont $45^{o}$}
  \put(129.07,17.22){\fontsize{14.23}{17.07}\selectfont $P'$}
  \put(232.94,17.22){\fontsize{14.23}{17.07}\selectfont $P$}
  \put(138.95,77.28){\fontsize{13.54}{15.24}\selectfont $\gamma'$}
  \put(236.04,77.16){\fontsize{13.54}{15.24}\selectfont $\gamma$}
  \end{picture}%
\else
  \setlength{\unitlength}{1bp}%
  \begin{picture}(336.87, 193.36)(0,0)
  \put(0,0){\includegraphics{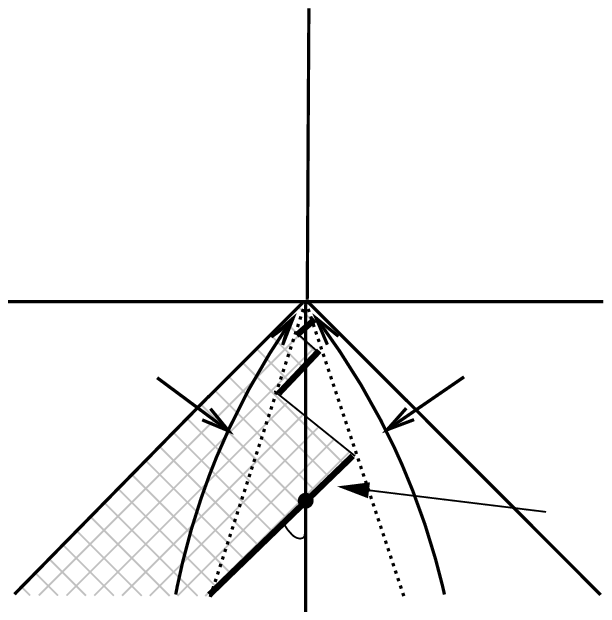}}
 \put(279.11,88.97){\fontsize{14.23}{17.07}\selectfont y}
 \put(196.23,176.58){\fontsize{14.23}{17.07}\selectfont t}
 \put(261.69,32.08){\fontsize{14.23}{17.07}\selectfont $L_{n}$}
 \put(170.56,8.73){\fontsize{14.23}{17.07}\selectfont $45^{o}$}
 \put(129.07,17.22){\fontsize{14.23}{17.07}\selectfont $P'$}
 \put(232.94,17.22){\fontsize{14.23}{17.07}\selectfont $P$}
\put(138.95,77.28){\fontsize{13.54}{15.24}\selectfont $\gamma'$}
  \put(236.04,77.16){\fontsize{13.54}{15.24}\selectfont $\gamma$}
  \end{picture}%
\fi\caption{\label{fig6}Slice of $M$ at $x=x_0$, as seen from the ``eye'' in
previous figure. The segments $L_{n}=\pi_{n}\cap \{x=x_0\}$ for
 some $x_0\in\R$ are represented. The curves $\gamma$ and $\gamma'$ which defines $P$ and $P'$ as IPs are here also depicted. In fact, $P=I^-(\gamma)$ includes  the region between the two half lines at $\pm 45$ degrees with respect to the axis, while $P'=I^{-}(\gamma')$ is only the subregion striped with thin lines at $\pm 45$ degrees.}
\end{figure}
\end{example}

\begin{example}\label{eluis} {\em
We are going to show that the limit operator $\lcrono$ of the
completion of a spacetime is not necessarily of first order (suggesting a procedure for the construction of examples of $k$-th order for any $k\in \N \cup \{\aleph_0\}$). To this aim, we explain first a key example which is described with further
details in \cite[Chapter 4]{A}.

The spacetime $M$ consists of $\LL^3$ with the following subsets
removed (see Figures \ref{noprimerorden}, \ref{firstorderarre}):
the causal future $J^{+}(0,0,0)$, the sheets $\{\Pi^{\infty}_{n}\}_{n\geq
1}$, where $\Pi^{\infty}_n=J^{-}(0,0,0)\cap \Pi_{n}$ and $\Pi_{n}$ is the plane $t-1/n=0$, the semi-discs $\{D_n\}_{n\geq 1}\cup D_{\infty}$ and
the sheets $\{\Pi^{l}_{n}\}_{l,n\geq 1}$ (obtained from
$\{\Pi^{\infty}_{n}\}_{n}$ by a convenient ``contraction and translation'').
The spacetime $M$ presents a number of distinguished
indecomposable sets, which are also depicted in the figures, say:
$P_{\infty}$, $P'_{\infty}$, $F=I^{+}(p_0)$ and $P_n, P'_n$ with $n\geq 1$; notice that each $(P_n,F)$ (rather than $p_{n}$) as well as $(P_{\infty},F)$ and $(P'_{\infty},\emptyset)$ are c-boundary points.

Consider the sequence $\sigma=\{x_n\}\subset M$, with
$x_n=(1/2,1/n,1/2)$ for all $n$. Then, reasoning as in Example \ref{ex1},
$\lcrono(\sigma)$ contains $\{(P_{n},F)\}_{n\geq 1}$ and $(P_{\infty},F)$. Moreover, $(P'_{\infty},\emptyset)
\in \lcrono(\{(P_{n},F)\})$. So, $\sigma $ converges to
$(P'_{\infty},\emptyset)$ with the chronological topology, as $(P'_{\infty},\emptyset)\in \lcrono^2(\sigma)$ (recall \eqref{aux2} and Proposition \ref{aux1}). Note,
however, that $P'_{\infty} \subsetneq P_{\infty} \subset LI(\{I^{-}(x_n)\}))$. 
Hence, $P'_{\infty}$ is not a maximal IP in $LS(\{I^{-}(x_n)\})$, and thus,
$(P'_{\infty},\emptyset) \not \in \lcrono(\sigma)$. In conclusion, $\lcrono$
is not of first order. However, it seems that all the possible limits are contained in $\lcrono^2$ (and making a straightforward modification of the example it would be contained in some $\lcrono^{k}$), so it is conceivable the existence of examples where $\lcrono$ is a $k$-th order limit operator for any $k\in \N \cup \{\aleph_0\}$.

 \begin{figure}
 \centering
 \ifpdf
   \setlength{\unitlength}{1bp}%
   \begin{picture}(523.80, 298.25)(0,0)
   \put(0,0){\includegraphics{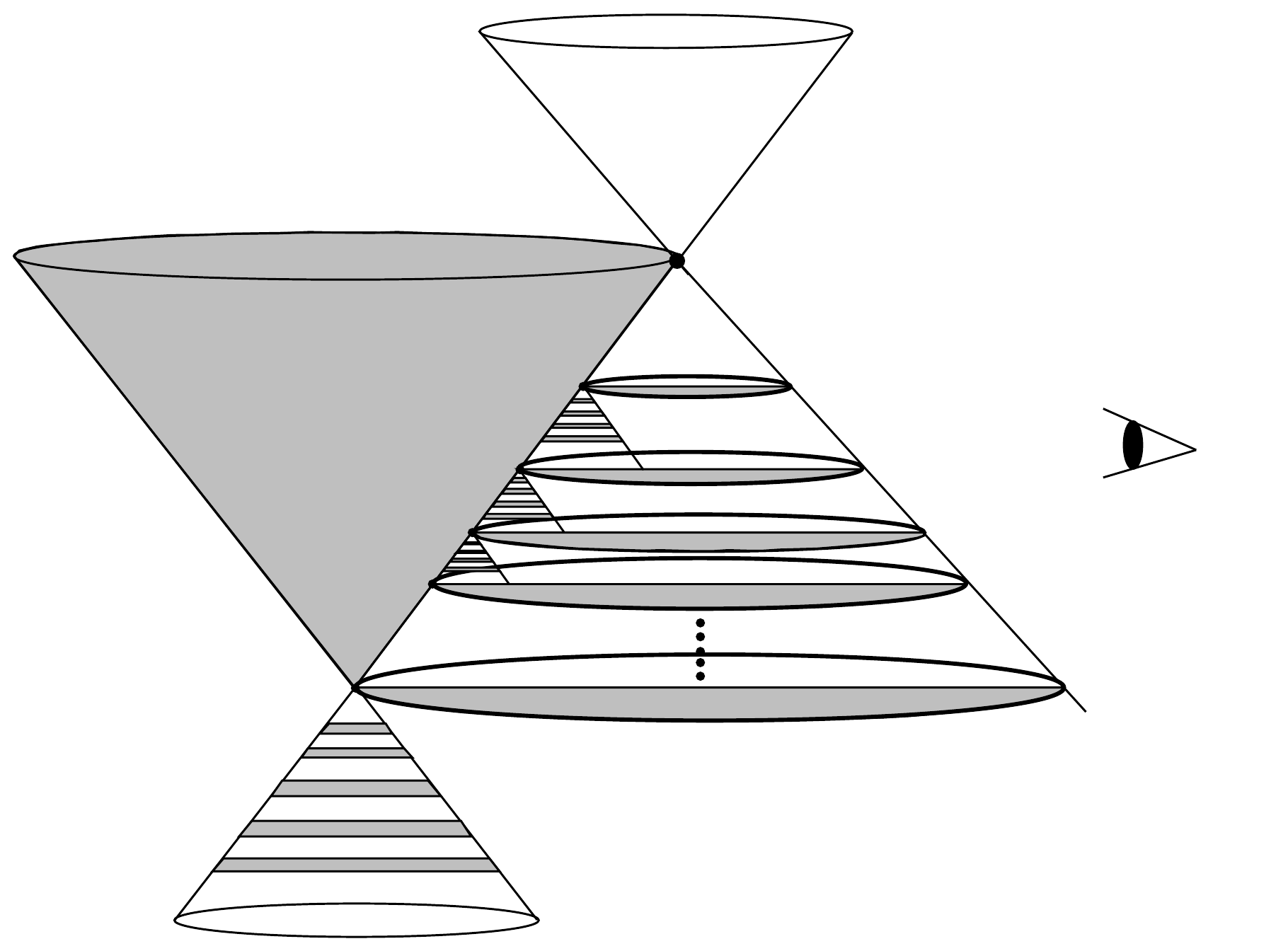}}
   \put(282.95,280.93){\fontsize{11.82}{14.18}\selectfont $p_0=(\frac{1}{2},0,\frac{1}{2})$}
   \put(217.13,230.41){\fontsize{11.82}{14.18}\selectfont $p_1$}
   \put(196.22,195.64){\fontsize{11.82}{14.18}\selectfont $p_2$}
   \put(175.42,170.89){\fontsize{11.82}{14.18}\selectfont $p_3$}
   \put(157.29,149.79){\fontsize{11.82}{14.18}\selectfont $p_4$}
   \put(70.90,102.80){\fontsize{11.82}{14.18}\selectfont $p_{\infty}=(0,0,0)$}
   \put(322.62,228.69){\fontsize{15.73}{18.87}\selectfont $D_1$}
   \put(352.04,196.33){\fontsize{15.73}{18.87}\selectfont $D_2$}
   \put(376.42,169.43){\fontsize{15.73}{18.87}\selectfont $D_3$}
   \put(394.49,148.84){\fontsize{15.73}{18.87}\selectfont $D_4$}
   \put(434.29,108.79){\fontsize{15.73}{18.87}\selectfont $D_{\infty}$}
   \put(101.31,9.92){\fontsize{14.23}{17.07}\selectfont $P_{\infty}$}
\put(151.31,22.92){\fontsize{14.23}{17.07}\selectfont $P'_{\infty}$}
   \put(253.61,209.44){\fontsize{11.38}{13.66}\selectfont $P_1$}
   \put(222.00,180.32){\fontsize{8.54}{10.24}\selectfont $P_2$}
   \put(263.53,331.83){\fontsize{15.73}{18.87}\selectfont $F$}
   \end{picture}%
 \else
   \setlength{\unitlength}{1bp}%
   \begin{picture}(523.80, 298.25)(0,0)
   \put(0,0){\includegraphics{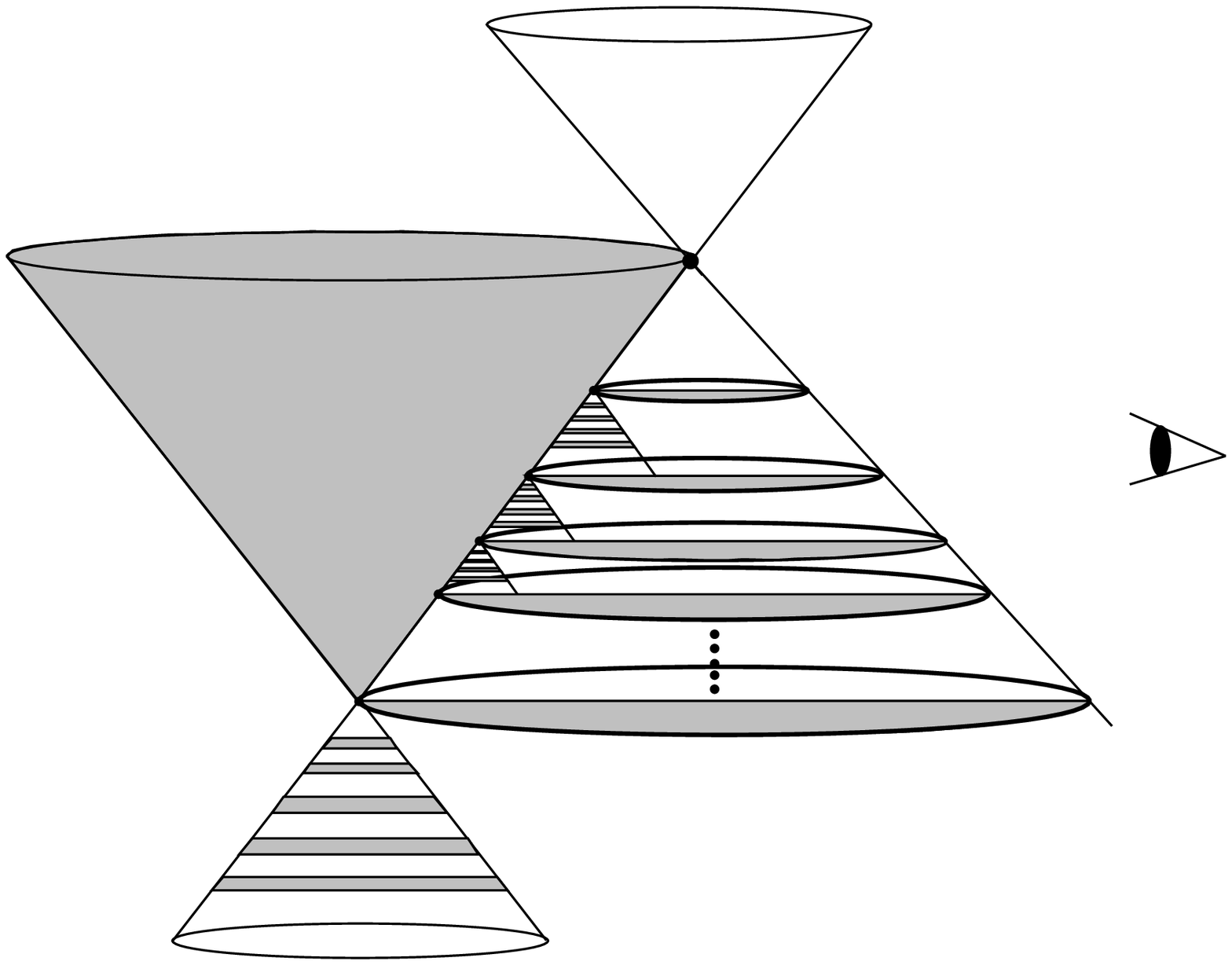}}
   \put(282.95,280.93){\fontsize{11.82}{14.18}\selectfont $p_0$}
   \put(217.13,230.41){\fontsize{11.82}{14.18}\selectfont $p_1$}
   \put(196.22,195.64){\fontsize{11.82}{14.18}\selectfont $p_2$}
   \put(175.42,170.89){\fontsize{11.82}{14.18}\selectfont $p_3$}
   \put(157.29,149.79){\fontsize{11.82}{14.18}\selectfont $p_4$}
   \put(124.90,107.80){\fontsize{11.82}{14.18}\selectfont $p_{\infty}$}
   \put(322.62,228.69){\fontsize{15.73}{18.87}\selectfont $D_1$}
   \put(352.04,196.33){\fontsize{15.73}{18.87}\selectfont $D_2$}
   \put(376.42,169.43){\fontsize{15.73}{18.87}\selectfont $D_3$}
   \put(394.49,148.84){\fontsize{15.73}{18.87}\selectfont $D_4$}
   \put(434.29,108.79){\fontsize{15.73}{18.87}\selectfont $D_{\infty}$}
   \put(131.31,18.92){\fontsize{14.23}{17.07}\selectfont $P_{\infty}$}
   \put(253.61,209.44){\fontsize{11.38}{13.66}\selectfont $P_1$}
   \put(222.00,180.32){\fontsize{8.54}{10.24}\selectfont $P_2$}
   \put(263.53,331.83){\fontsize{15.73}{18.87}\selectfont $F$}
   \end{picture}%
\fi
\caption{Representation of the spacetime $M$, obtained by removing
from $\LL^{3}$ the grey regions.
%: $J^{+}(0,0,0)$, the plaques
%$\{\pi'_{m}\}_{m\geq 2}$ with $\pi'_m=J^{-}(0,0,0)\cap \pi_{m}$
%and $\pi_{m}\equiv t-1/m=0$, the semi-discs $\{D_m\}_m$ and
%$\{\pi_{m,l}\}_{m,l\geq 2}$.
The figure also shows the distinguished IPs, $P_\infty$, $P'_{\infty}$ and
$\{P_n\}_{n\geq 1}$ (see Figure \ref{firstorderarre}). Observe that every $P_n$ and $P'_n$ consists of a
``contraction plus a translation'' of $P_{\infty}$ and $P'_{\infty}$ resp.} \label{noprimerorden}
\end{figure}

\begin{figure}
\centering
\ifpdf
  \setlength{\unitlength}{1bp}%
  \begin{picture}(300.29, 309.05)(0,0)
  \put(0,0){\includegraphics{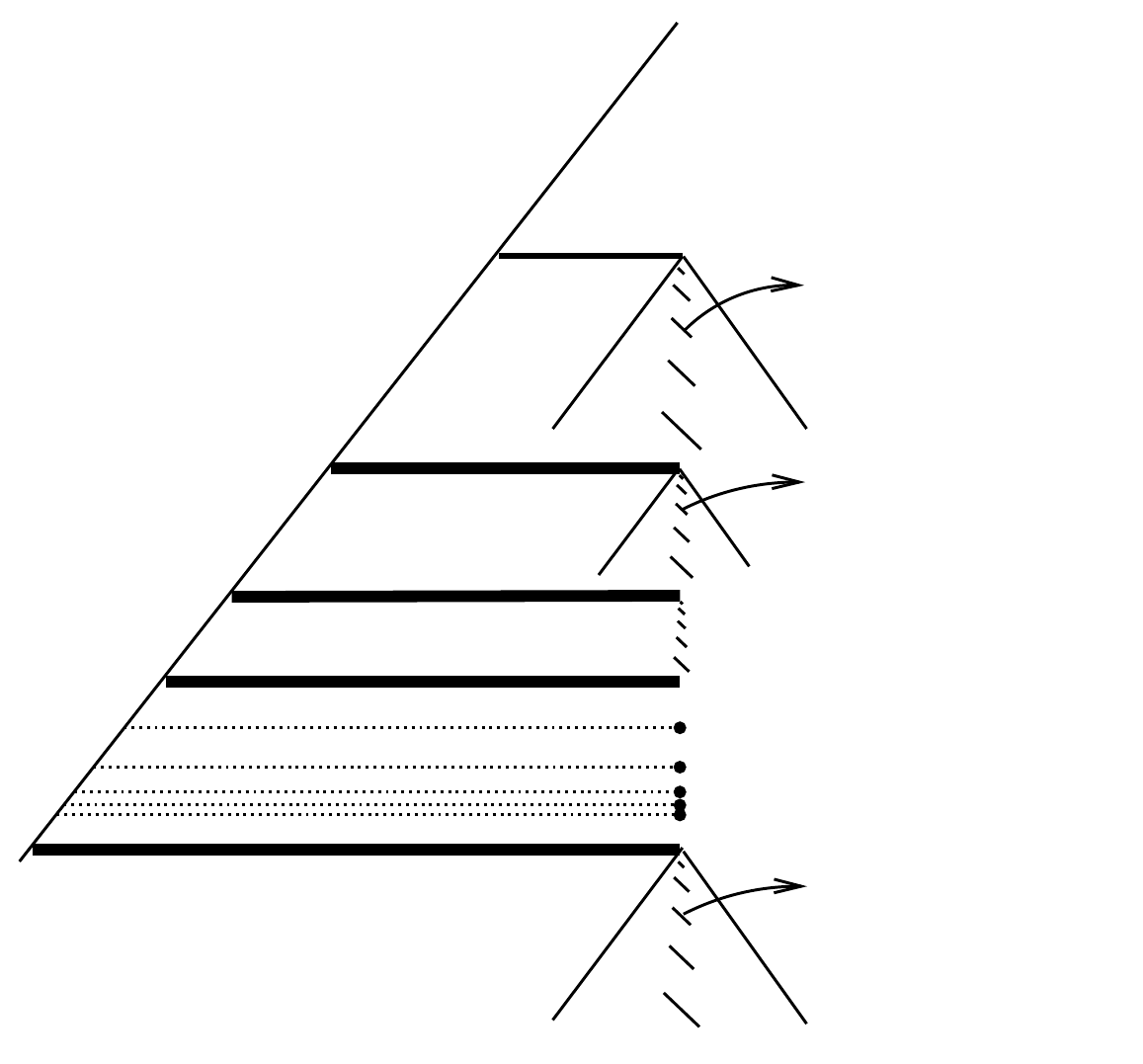}}
  \put(175.54,187.51){\fontsize{13.98}{16.78}\selectfont $P_1$}
    \put(207.05,187.51){\fontsize{13.98}{16.78}\selectfont $P'_1$}
    \put(238.55,224.81){\fontsize{13.98}{16.78}\selectfont $\Pi_n^1$}
    \put(155.51,239.86){\fontsize{14.30}{17.16}\selectfont $D_1$}
    \put(114.42,179.03){\fontsize{14.30}{17.16}\selectfont $D_2$}
    \put(183.92,145.53){\fontsize{9.55}{11.46}\selectfont $P_2$}
    \put(200.85,145.53){\fontsize{9.55}{11.46}\selectfont $P'_2$}
    \put(238.55,167.13){\fontsize{13.98}{16.78}\selectfont $\Pi_n^2$}
    \put(175.54,9.22){\fontsize{13.98}{16.78}\selectfont $P_\infty$}
    \put(207.05,8.68){\fontsize{13.98}{16.78}\selectfont $P'_{\infty}$}
    \put(74.32,47.78){\fontsize{14.30}{17.16}\selectfont $D_{\infty}$}
    \put(234.55,49.29){\fontsize{13.98}{16.78}\selectfont $\Pi_n^{\infty}$}

  \end{picture}%
\else
  \setlength{\unitlength}{1bp}%
  \begin{picture}(300.29, 309.05)(0,0)
  \put(0,0){\includegraphics{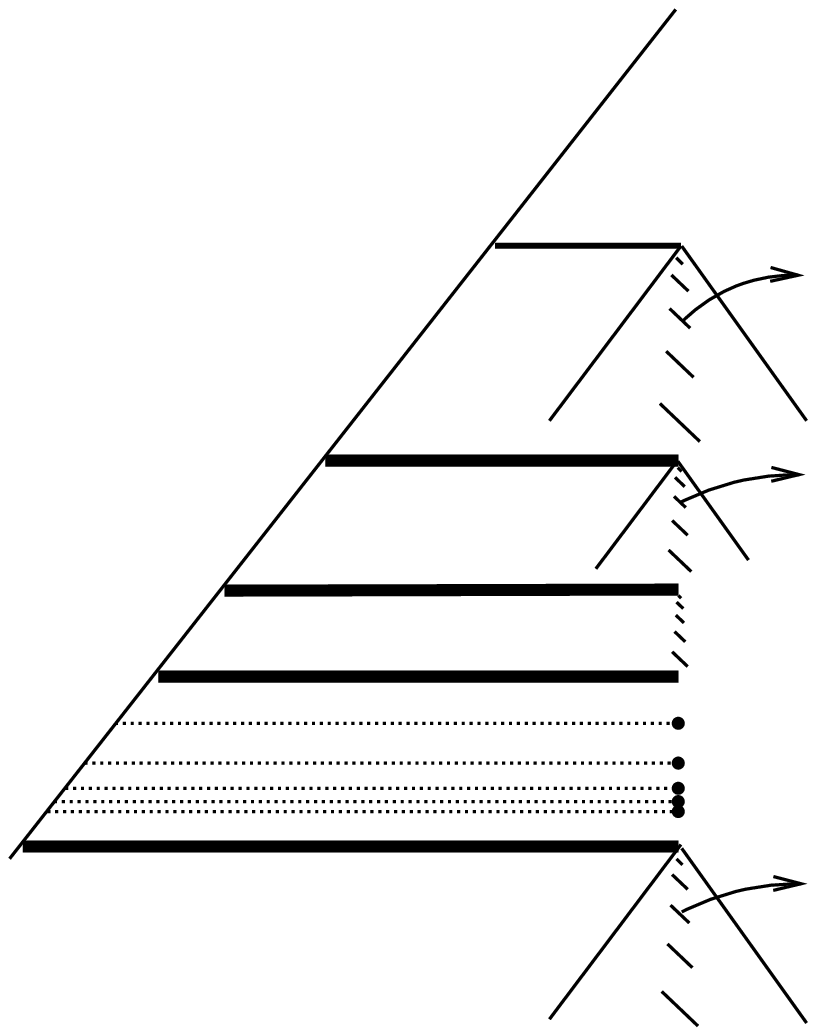}}
  \put(179.54,187.51){\fontsize{13.98}{16.78}\selectfont $P_1$}
  \put(207.05,187.51){\fontsize{13.98}{16.78}\selectfont $P'_1$}
  \put(238.55,224.81){\fontsize{13.98}{16.78}\selectfont $\Pi_n^1$}
  \put(155.51,239.86){\fontsize{14.30}{17.16}\selectfont $D_1$}
  \put(114.42,179.03){\fontsize{14.30}{17.16}\selectfont $D_2$}
  \put(180.92,145.53){\fontsize{9.55}{11.46}\selectfont $P_2$}
  \put(200.85,145.53){\fontsize{9.55}{11.46}\selectfont $P'_2$}
  \put(234.55,167.13){\fontsize{13.98}{16.78}\selectfont $\Pi_n^2$}
  \put(166.92,9.22){\fontsize{13.98}{16.78}\selectfont $P_\infty$}
  \put(199.31,8.68){\fontsize{13.98}{16.78}\selectfont $P'_{\infty}$}
  \put(74.32,47.78){\fontsize{14.30}{17.16}\selectfont $D_{\infty}$}
  \put(234.55,49.29){\fontsize{13.98}{16.78}\selectfont $\Pi_n^{\infty}$}
  \end{picture}%
\fi
\caption{Section of $M$ as seen from the ``eye'' in previous
figure. Note that, between two semi-discs $D_i$ and $D_{i+1}$, we have the same effect as in Example \ref{ex1} (compare with Figures \ref{fig5} and \ref{fig6}). Moreover, for all $n$, $P_{\infty}\not\subset P_{n}$ (in
fact, $P_{m}\not\subset P_{n}$ for $m>n$), but $P'_{\infty}\subset P_n$. This implies $P'_{\infty} \subset LI(\{P_{n}\})$ and
$P'_{\infty}$ is maximal IP in $LS(\{P_{n}\})$, that is, $(P'_{\infty},\emptyset)
\in \lcrono(\{(P_{n},F)\})$.} \label{firstorderarre}
\end{figure}
}
\end{example}

\begin{example}\label{ex0}{\em
Here, we present an example clarifying our definition of condition (\Asep). Concretely, we will justify the requirement that the open set $U$ belongs to the original topology $\tau_{M}$ of the manifold $M$ and not to the topology to be obtained. Basically, the reason of our choice is that the alternative topology might not preserve the original topology of the manifold $M$, as the following example shows.
%we will provide an explicit example of a topology, a priori as good as $\tnew$, which does not satisfy previous requirement, and thus, it does not preserve the original topology of the manifold.

Consider the spacetime $M$ depicted in Figure \ref{fig5}. It is not difficult to check that the
chronological limit operator $\lcrono$ is of first
order. Let us define a new limit operator $\lnewm$ as
follows (compare with (\ref{def2})):
\[
\lnewm(\sigma):=\left\{\begin{array}{lc}\lcrono(\sigma)\cap
\partial M & \hbox{if $\exists$
$p\in M$ and $(P,F)\in\partial M$ such that $p,(P,F)\in
\lcrono(\sigma)$}  \\ \lcrono(\sigma) & \hbox{otherwise,}
\end{array}\right.
\]
\noindent and denote by $\tnewm$ the topology associated to this limit
operator. Since $\lcrono$ is of
first order, it follows, by the same arguments as in Section \ref{sec3.3}, that $\lnewm$ is a first order limit operator (recall Remark \ref{g}). Moreover, its associated topology $\tnewm$ is finer than
$\tcrono$, T$_2$-separates the points of the boundary from
the points of the manifold, and it is minimally finer among the
sequential ones satisfying such properties. However, the sequence
$\sigma=\{p_n\}$ in Figure \ref{fig5} does not converge to $p$ with $\tnewm$, since it satisfies
$p,(P',\emptyset)\in \lcrono(\sigma)$ and so, $p\notin \lnewm(\sigma)$. Taking into account that $\sigma$ clearly
converges to $p$ with the manifold topology, we deduce that $\tnewm$ does not
preserves the manifold topology on $M$.}
\end{example}

\section*{Acknowledgments}

The authors are partially supported by the Spanish
Grants MTM2013-47828-C2-1-P and MTM2013-47828-C2-2-P (MINECO and FEDER funds).
%The first and third-named authors are also partially supported by the Spanish Grants MTM2013-47828-C2-2-P and MTM2013-47828-C2-1-P, resp.
The second-named author is also supported
by FAPESP (Funda\c{c}\~ao de Amparo \'{a} Pesquisa do Estado de S\~ao Paulo, Brazil) Process 2012/11950-7.

\end{document}